\RequirePackage{fix-cm}
\documentclass[]{svjour3}       
\pdfoutput=1

\usepackage[fleqn]{amsmath}
\usepackage{epsfig}
\usepackage{graphicx}
\usepackage{capt-of}
\usepackage{amssymb}
\usepackage{natbib}
\usepackage{abrev-jour-long}

\usepackage[normalem]{ulem}
\usepackage[hidelinks]{hyperref}
\usepackage{doi}

\usepackage{color}
\usepackage[dvipsnames]{xcolor}

\bibpunct{(}{)}{;}{a}{}{,}

\smartqed  
\begin{document}

\title
{Integrability of motion around galactic razor-thin disks}


\author
{Ronaldo S. S. Vieira \and Javier Ramos-Caro}



\institute{Ronaldo S. S. Vieira \at
	    		Instituto de Astronomia, Geof\'{i}sica e Ci\^encias
			Atmosf\'{e}ricas, Universidade de S\~{a}o Paulo, \mbox{05508-090},
			S\~{a}o Paulo, SP, Brazil \\
              \email{rss.vieira@usp.br} 
           \and
           Javier Ramos-Caro \at
              Departamento de F\'isica, Universidade
	      Federal de S\~{a}o Carlos, 13565-905, SP, Brazil\\
	      \email{javier@ufscar.br}
}

\date{Received: date / Accepted: date}

\maketitle

\begin{abstract}
We consider the three-dimensional bounded motion of a test particle around
razor-thin disk configurations, by focusing on  the adiabatic invariance of the   vertical action associated with
disk-crossing orbits.
We find that it leads to an approximate third integral of motion
predicting envelopes of the form $Z(R)\propto[\Sigma(R)]^{-1/3}$,
where $R$ is the radial galactocentric coordinate, $Z$
is the z-amplitude (vertical amplitude) of the orbit and $\Sigma$ represents the surface mass density of the thin disk.
This third integral, which was previously formulated for
 the case of flattened 3D configurations, is tested for a variety of trajectories in different thin-disk models.
\end{abstract}

\section{Introduction}\label{sec:intro}

Many galaxies in the Universe are nearly axisymmetric, with a
mass distribution formed by several components: a thin disk, a central bulge and a
surrounding halo. In consequence, there is a number of mass models incorporating one or all of these
features, depending on the particular situation \citep[]{freeman1970ApJ, kent1986AJ, kent1987AJ, sofue-honma-omodaka2009PASJ}.
But, in all cases, the disk component  represents  a significant
percentage of the galactic mass, taking into account that the main part of
the stellar population is located there. For systems in which	 this  component is highly flattened, we can
use thin disks in the limit of negligible thickness (razor-thin disks) in order to provide preliminary,
but simple and tractable, models
for the stellar mass distribution.
In consequence, the formulation and characterization of thin-disk  models have been, for decades,
an issue of interest in galactic dynamics
\citep[see for example][as well as
\citealt{binney-tremaine:GD} and references therein]{hunter1963MNRAS, morgan-morgan1969PR,hunter2003LNP,hunter2005NYASA,
gonzalez-reina2006MNRAS, pedraza-ramoscaro-gonzalez2008MNRAS}.

Razor-thin disks (RTDs) have also been used in other branches of astrophysics.
These models can be used to describe self-gravitating disks around a
central black hole
\citep[see][]{lemos-letelier1994PRD,semerak-sukovaMNRAS2010,loraclavijo-ospinahenao-pedraza2010PRD}
and self-gravitating rings \citep{letelier2007MNRAS, vogt-letelier2009MNRAS,iorio2012EMP}.
The last issue was partially encompassed in \citet{ramoscaro-pedraza-letelier2011MNRAS}, for instance,
where a linear stability study of the monopole-ring system was performed by using superpositions of Morgan \& Morgan disks.
Similar studies were conducted by the authors in the Newtonian realm of galactic dynamics
(\citealt{ramoscaro-lopezsuspes-gonzalez2008MNRAS}; \citealt{pedraza-ramoscaro-gonzalez2008MNRAS}).

The analysis of the motion
of test particles in the gravitational field generated by these axially symmetric configurations
 is a fundamental issue, which helps
us to characterize the dynamics
in a given model and establish
predictions in situations where they can be applied.
For example, the study of orbits in the equatorial
plane can improve our understanding about
the dynamics of intragalactic stellar motion or the flow of particles
in accretion disks around black holes. On the other hand, the integrals of motion
associated with three-dimensional orbits are a fundamental key to find
the galaxy's distribution function in stationary state \citep{binney-tremaine:GD}. They set the basis for the
formulation of dynamical models for the Galaxy
\citep*[]{dezeeuw1987IAUS,binney2010MNRAS,binneyMcmillan2011MNRAS,binney2012MNRAS,binneySanders2014IAUS}.

In this paper
we  study  three-dimensional bounded motion of
test particles around razor-thin distributions of matter, by focusing on
the problem of the so-called disk-crossing orbits, i.e. trajectories which cross back
and forth through the stellar disk.
Numerical experiments reveal prominent regions of KAM curves
in the corresponding phase space, suggesting the
existence of a third (non-classical) integral of motion
\citep[]{saa-venegeroles1999PhLA,hunter2003LNP, hunter2005NYASA,ramoscaro-lopezsuspes-gonzalez2008MNRAS}.
This is an interesting fact, taking into account that  smooth versions
of KAM theorem cannot be applied to such distributions, due to the discontinuity
introduced by the razor-thin layer.
The principal motivation of this work is to provide a tractable
expression for the non-analytical integral of motion governing disk-crossing particles.
The results we will show here, suitable for situations with discontinuous gravitational fields,
can be considered as a complement  to the analysis conducted in
\cite{vieira-ramoscaro2014ApJ}, applicable only to continuous fields due to smooth matter
distributions.

The problem of finding an analytical expression for the third integral of motion
governing the orbits of stars in a galaxy has been calling the attention of the astrophysics community for decades
\citep{contopoulos1960ZA,contopoulos1963AJ,
contopoulos2001development,contopoulosOCDA2002,
hieratinta1987PhR,bienaymeTraven2013AA,bienaymeRobinFamaey2015AA}.
This issue can be addressed in a simple way for some elementary situations, as for example, the description of
 quasi-circular orbits around
 the equatorial plane
of a three-dimensional flattened distribution. In
such a case, the epicyclic approximation holds and, by invoking the adiabatic invariance of the corresponding vertical action,
 one can construct an approximate formula for the third integral, $I_3$, in cylindrical coordinates ($R,\varphi,z$), evaluated at the orbit's envelope $Z(R)$ \cite*[]{binney-tremaine:GD}:
 \begin{equation}\label{I3-tremaine}
    I_3= Z(R) \Phi_{zz}^{1/4}(R,0),
 \end{equation}
where $Z(R)$ is the orbit's vertical amplitude at a radial coordinate $R$, and $\Phi_{zz}$
represents the second derivative  of the gravitational potential with respect to $z$.
Unfortunately, the range of validity of such a formula is very limited and, more important, it is not applicable to potentials generated by
RTDs, for which $\Phi_{zz}$ cannot be defined inside the matter distribution.

We present in this paper a new approximate third integral of motion, valid for disk-crossing orbits
in a potential due to a RTD (a different approximate third integral for a class of $C^0$ systems was presented in \citealp{varvoglis1985JPhys}).
This integral describes the shape of nearly equatorial orbits in terms of $\Sigma$, the surface matter density of the razor-thin layer.
The approach begins with a presentation of the stability analysis of equatorial circular orbits
belonging to the razor-thin distribution of matter, in order to establish  stability criteria appropriated to situations involving
field discontinuities.
The results of  this analysis are incorporated in the description of
adiabatic invariants, obtaining an expression as simple as (\ref{I3-tremaine}) when evaluated at the orbit's envelope:
\begin{equation}\label{I3-RJ}
    I_3= Z(R)\Sigma^{1/3}(R),
 \end{equation}
which means that the vertical amplitude $Z$, at a galactocentric radius $R$, is determined by the
surface mass distribution $\Sigma$ through a relation of the form \mbox{$Z\propto\Sigma^{-1/3}$}, setting
the shape of the envelope traced in the configuration space $(R,z)$ (also called \textit{meridional plane}).
Remarkably, the envelopes
of nearly equatorial disk-crossing  orbits are determined
exclusively by the surface density of the razor-thin disk, even when there are other matter sources around it (halo, thick disk, etc).

In the following, we present the procedure to obtain formula (\ref{I3-RJ}), which represents
the RTD-version of formula (\ref{I3-tremaine}).
At first, we present a brief revision on the formulation of models including RTDs (sec. \ref{sec:thin}), in order to
set the properties of the gravitational potential determining the test-particle motion, 
which will be addressed in sec. \ref{sec:epicyclic}.
There, the main problem will be to study
the stability of equatorial circular orbits under perturbations in $R$- (radial) and $z$- (vertical) directions, in
order to obtain stability criteria suitable  to regions belonging to the thin layer.
Then, we incorporate such criteria
in the description of disk-crossing orbits. In particular, we show that a variety of these orbits can be described
by an approximate third integral of motion given by Eq. (\ref{I3-RJ})
(sec. \ref{sec:integrability}). We test our results with numerical experiments in different razor-thin disk models (sec. \ref{sec:numerical}). Finally, we  present our conclusions in sec. \ref{sec:conclusions}.

\section{Galactic Models via RTDs}\label{sec:thin}

As it was pointed out in the Introduction, many galaxies are modeled as
 an axisymmetric RTD surrounded by an axisymmetric 3D smooth density distribution, which is symmetric with respect to the
equatorial plane. The total density distribution can be written as
  \begin{equation}\label{totaldensity}
   \rho(R,z) = \Sigma(R)\delta(z) + \rho_s(R, z),
  \end{equation}
where $\delta$ is the Dirac delta, $\Sigma$ is the surface density distribution of the RTD and
$\rho_s$ describes the surrounding matter. The
gravitational potential of the system is
  \begin{equation}\label{totalpotential}
   \Phi(R, z) = \Phi_\Sigma(R,z) + \Phi_s(R, z),
  \end{equation}
where $\Phi_\Sigma$ and $\Phi_s$ are the contributions due to $\Sigma$ and $\rho_s$, respectively, in
such a way that they obey the field equations
$\nabla^2\Phi_\Sigma = 4\pi G \Sigma(R)\delta(z)$ and
$\nabla^2\Phi_s = 4\pi G \rho_s$.

The assumption  that the distribution
 $\rho_s$ is symmetric with respect to the equatorial plane  implies the same reflection symmetry for the gravitational potential,
i.e. $\Phi(R, z) = \Phi(R, -z)$. Then,
the $z$-dependence of $\Phi$ is solely on $|z|$
and, from Poisson's equation, $\Sigma$ is given by
  \begin{equation}
   \Sigma(R) = \frac{1}{2\pi G}\frac{\partial \Phi}{\partial |z|}\bigg|_{z = 0},\label{sigma}
  \end{equation}
which is equivalent to the expression appearing in \citet{gonzalez-reina2006MNRAS} and references therein.

\section{Stability of equatorial circular orbits in RTD models}\label{sec:epicyclic}

The motion of test particles under the action of $\Phi(R,z)$ can be described
by the Hamiltonian
$H = (P_R^2 + P_z^2)/2  + \Phi_{eff}$ (per unit mass),
where $P_R=\dot{R}$, $P_z=\dot{z}$ and $\Phi_{eff}= \Phi(R, z) + \ell^2/(2R^2)$ is the effective potential, determined
 by $\ell$, the $z$-component of the particle's specific angular momentum.
Consider an equatorial circular orbit of radius $R$ under the
action of a small vertical perturbation\footnote{Here, the term ``small'' means
\textit{small enough to neglect variations in the projection of the orbit on
the $z=0$ plane}.}. This perturbation can be seen as an instantaneous vertical increase $v_{0z}$ in the
velocity of the particle, which does not affect the value of $\ell$.
In order to study the evolution of the perturbation in the course of time, we have to use the
$z$-equation of motion, $\ddot z = -\partial\Phi/\partial z$, which can be written as
  \begin{equation}
   \ddot{z} = - [2 \Theta(z)-1]\frac{\partial \Phi}{\partial |z|},
  \end{equation}
where $\Theta$ is the Heaviside function. From the above expression it can be noticed that the  vertical
perturbation will remain small if and only if  $[\partial \Phi/\partial |z|]_{z = 0} > 0$.
It guarantees that,
if $v_{0z}$ is small enough, the perturbed trajectory will
oscillate around the original one for long times.
This assertion can be seen, for instance, from the expansion of $\Phi_{eff}(R,z)$
in powers of $|z|$:
  \begin{equation}\label{taylor1}
   \Phi_{eff}(R,z) = \Phi_{eff}(R,0) + \frac{\partial \Phi}{\partial |z|}(R,0)\, |z| + \mathcal{O}(|z|^2).
  \end{equation}
The particle will always cross the disk with a velocity whose vertical component
has absolute value $|v_{0z}|$.
 This is a consequence of: (i) the conservation of mechanical energy;
(ii) the assumption that the $R$-coordinate does not change in the process and
 (iii) the fact that the discontinuity is only in acceleration. So, just after crossing the disk, the particle will also
have a velocity with vertical component of magnitude $|v_{0z}|$, which means that motion after crossing the disk will have the same
behavior as motion before crossing it. Therefore, the perturbed orbit remains oscillatory around
the original one for sufficiently small initial vertical velocity $v_{0z}$.

According to (\ref{sigma}), condition $[\partial \Phi/\partial |z|]_{z = 0} > 0$ can be written in terms of the
surface mass density $\Sigma$.
Then we can state that in a Newtonian razor-thin disk model, a necessary and sufficient condition for vertical stability of a circular
orbit of radius $R$ is $\Sigma(R) > 0$. But we know that all the disk models, intended to represent realistic
matter distributions, satisfy such a condition.
Then we can state that  \textit{the circular equatorial orbits in RTD models are always stable under small vertical
perturbations}.

By introducing (\ref{taylor1}) in the equations of motion, it is possible to estimate, for a
sufficiently small vertical perturbation, the characteristic period $T$ and amplitude $Z$ of the corresponding oscillations around the equatorial plane.
For the period we obtain $T = 2 v_{0z}/(\pi G \Sigma(R))$,
whereas for the amplitude of the oscillation, we can write $ Z= v_{0z}^2/(4 \pi G \Sigma(R))$.
As expected, the amplitude of oscillations is inversely proportional to the surface mass density
(but not to the volumetric mass density
of the surrounding matter, if present).

As it is well known, radial stability is guaranteed once Rayleigh's criterion is satisfied \citep{landauLifshitzFM,letelier2003PRD}.

\section{Approximate integrability of motion near a stable circular orbit}\label{sec:integrability}

We remarked in the Introduction that motion is integrable near a stable circular orbit,
a well-known result in the case of smooth density distributions.
There are many evidences of this general behavior for razor-thin disks
in the literature
\citep{saa-venegeroles1999PhLA, hunter2003LNP, hunter2005NYASA, pedraza-ramoscaro-gonzalez2008MNRAS,ramoscaro-lopezsuspes-gonzalez2008MNRAS,
gonzalez-plataplata-ramoscaro2010MNRAS,ramoscaro-pedraza-letelier2011MNRAS},
where it is found numerically, by means of Poincar\'e sections, that motion is integrable around what appears to be a stable point
of the effective potential -- corresponding to a stable circular orbit in the equatorial plane. These evidences also show that
the integrable domain goes well beyond the neighborhood of the stable circular orbit, reaching regions where the approximation of a separable
potential is not valid anymore.

The integrability of motion near the stable circular orbit for density profiles of the form (\ref{totaldensity})
follows from the separability of the
effective potential near the stable point $(R_{o},0)$. In fact, up to first order in $|z|$, we have
  \begin{eqnarray}
   \Phi_{eff}(R,z) &\approx& \Phi_{eff}(R,0) + \frac{\partial\Phi_{eff}}{\partial |z|}(R,0) |z|\nonumber \\
    &=& \Phi_{eff}(R,0) + 2\pi G\Sigma(R) |z|.\label{potaprox}
  \end{eqnarray}
Since the orbit is radially stable, we can approximate $\Sigma(R)\approx\Sigma(R_o)$, discarding higher-order terms. Thus,
  \begin{equation}\label{Aphiapprox}
   \Phi_{eff}(R,z) \approx \Phi_{eff}(R,0) + 2\pi G\Sigma(R_o) |z|,
  \end{equation}
and the corresponding approximate Hamiltonian, $H = (P_R^2 + P_z^2)/2 + \Phi_{eff}$,
is separable.
In this way, we obtain by quadratures two independent approximate  integrals of motion, namely the approximate action variables, near $(R_{o},0)$: one
for the $R$-coordinate, $J_{R}$, and another for the $z$-coordinate, $J_{z}$, which is of special interest here.

It can be shown that the vertical action $J_{z}$ can be written as
\begin{equation}\label{Jz}
   J_z = \frac{4\sqrt{2}}{3\pi}\left[2\pi G \Sigma(R_{o})\right]^{1/2}Z^{3/2}.
\end{equation}
This expression helps us to study the effects of adiabatic variations in the approximate potential of Eq. (\ref{Aphiapprox}).
Following the procedure presented in
section 3.6 of \citet{binney-tremaine:GD}, we obtain
an expression that relates the $z$-amplitudes of a given orbit at different values of the radial coordinate ($R\neq\tilde{R}$):
  \begin{equation}\label{eq:ZZ'sigma}
   \frac{Z(R)}{Z(\tilde{R})} = \bigg[\frac{\Sigma(\tilde{R})}{\Sigma(R)}\bigg]^{1/3}.
  \end{equation}
This relation, which is the same as (\ref{I3-RJ}), determines the ``envelope'' of the orbit in the meridional plane.
It works well whenever the vertical oscillations are faster than radial oscillations, as in the case of orbits
which deviate only slightly from the equatorial plane.
 Equation (\ref{eq:ZZ'sigma}), valid for razor-thin disks (and previously presented in \citealp{vieiraRamoscaro2015MG13}), is the analogue of the corresponding
well-known relation for smooth potentials \citep[eq. (\ref{I3-tremaine}), which is the same as eq. 3.279 of][]{binney-tremaine:GD}.

Finally, we point out that the approximated third integral, associated with $J_{z}$, takes the form
\begin{equation}\label{eq:I3integral}
I_3= \frac{1}{2\pi G}\left[\Sigma(R)\right]^{-2/3}
\bigg[\frac{1}{2}P_z^2+ 2\pi G\Sigma(R) \,|z|\bigg]
\end{equation}
in terms of the phase-space coordinates,
which can be compared with other approximated expressions. It can be shown that the above relation
implies (\ref{eq:ZZ'sigma}) when evaluated along the orbit's envelope $P_z=0$, relating
the vertical amplitude to the radial coordinate.
The vertical action $J_z$ relates to $I_3$ by
  \begin{equation}\label{Jzintegral}
   J_z=\frac{8G^{1/2}}{3\pi^{1/2}} \big[I_3\big]^{3/2}
  \end{equation}
(see Eq. (\ref{Jz})), so its expression in terms of phase-space variables is obtained using Eq. (\ref{eq:I3integral}).

\section{Examples via numerical experiments}\label{sec:numerical}

In order to check the validity of (\ref{eq:ZZ'sigma}), we can  compare the envelopes that it describes with
the envelopes of numerically calculated off-equatorial orbits in a given potential $\Phi_{eff}$.
Here we perform numerical experiments for the motion
of test particles around mass distributions of the form (\ref{totaldensity}), by considering
 several razor-thin disk potentials.
These ``nearly circular'' orbits may be obtained by considering orbits
with relative energy much smaller than the binding
energy of the particle:
$\Delta E\equiv|E-\Phi_{eff}(R_o,0)|/|\Phi_{eff}(R_o,0)|\ll1$. It is expected that, in this situation, the predicted $Z(R)$
matches the real envelope, since for this energy range the orbits would be bounded to a region in the meridional plane where the approximations
made in the former section are reasonable. On the other hand, to check the validity of (\ref{eq:I3integral}), we compute the time series of
$I_{3}$ along the interval of integration of the orbits and compare the surfaces of section of numerically integrated
orbits with the phase-space curves associated with $I_{3}$.

We also analyzed a number of situations
in which the integrated orbits are far from being considered as  ``nearly equatorial'' or ``nearly circular'', so we could
estimate the range of validity of (\ref{eq:ZZ'sigma}). Apart from the fact that, as we increase the energy of the system,
the vertical amplitude of the corresponding orbit tends to grow (as well as its radial range), going beyond
the expected region of validity of (\ref{eq:ZZ'sigma}), there is also another phenomenon which would invalidate the predictions
of the approximate third integral: as energy grows, the nonlinearity of the Hamiltonian flow implies that the secondary resonance islands
in phase space also tend to increase in size.
Since the adiabatic approximation considers a separable potential, it is likely that this approximation would
break down near these islands (whose orbits in the meridional plane would deviate only slightly from the corresponding periodic orbits).
In fact, this is what we found in all the cases analyzed, as we describe next.

\begin{figure}
\epsfig{figure=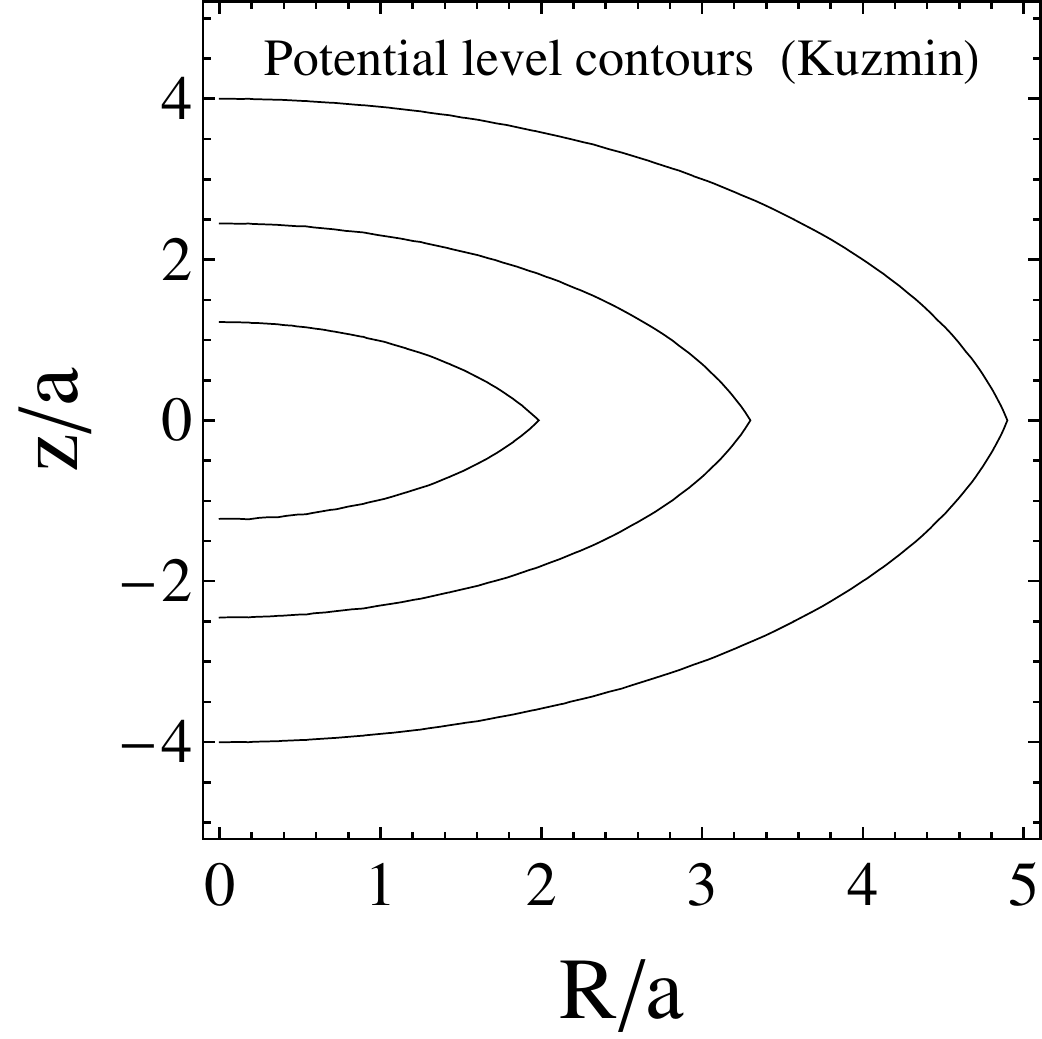,width=0.33\columnwidth ,angle=0}\quad
\epsfig{figure=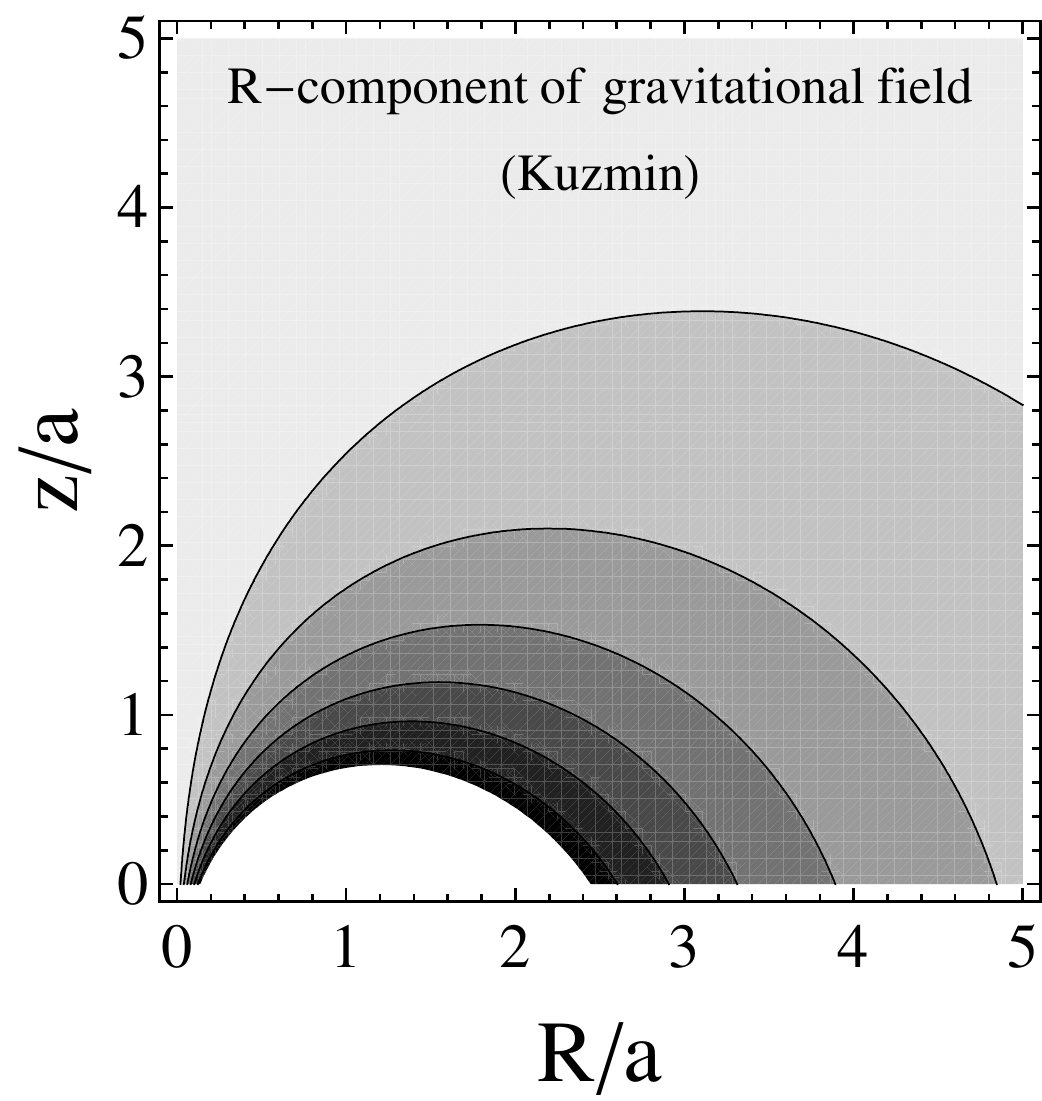,width=0.31\columnwidth ,angle=0}\quad
\epsfig{figure=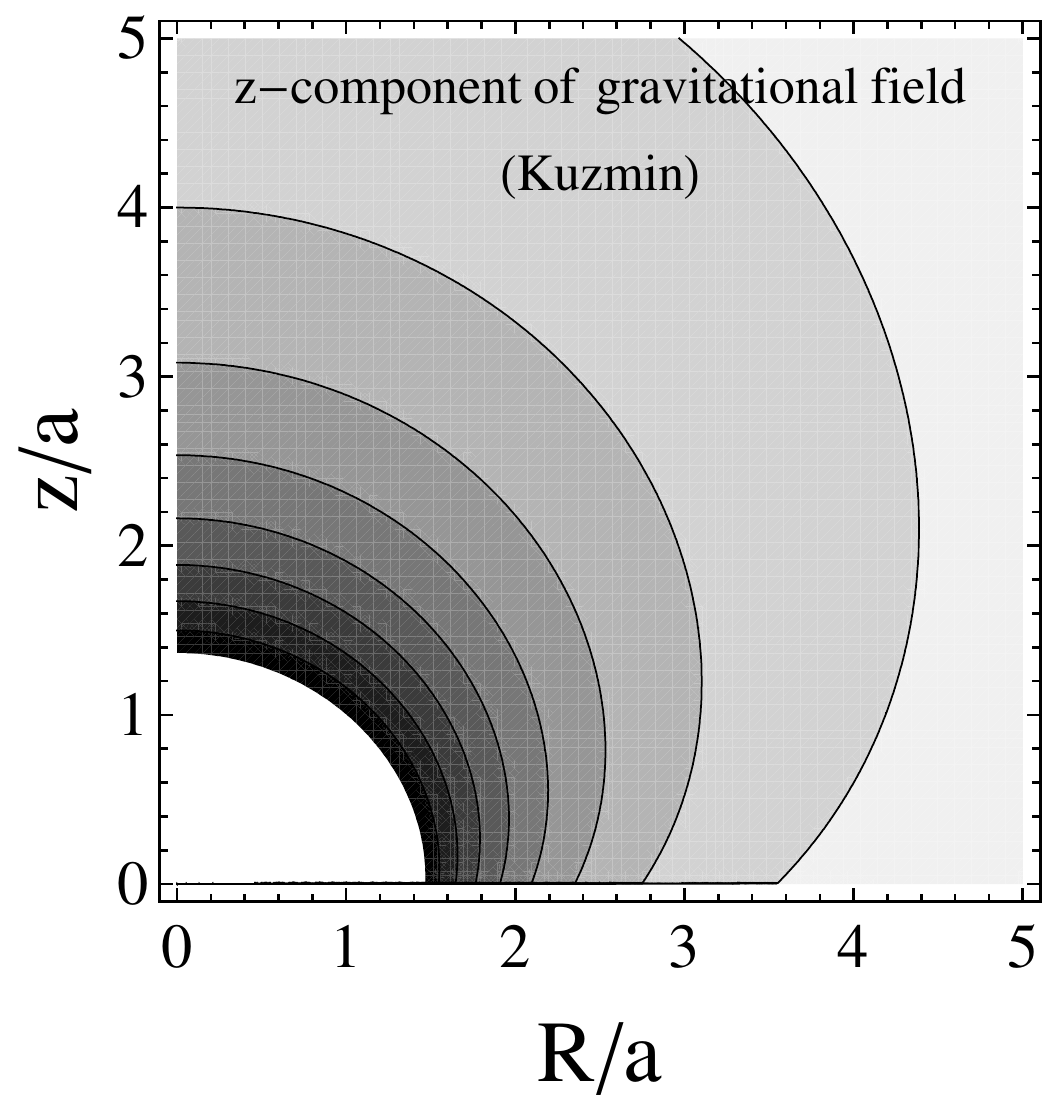,width=0.31\columnwidth ,angle=0}\quad
\epsfig{figure=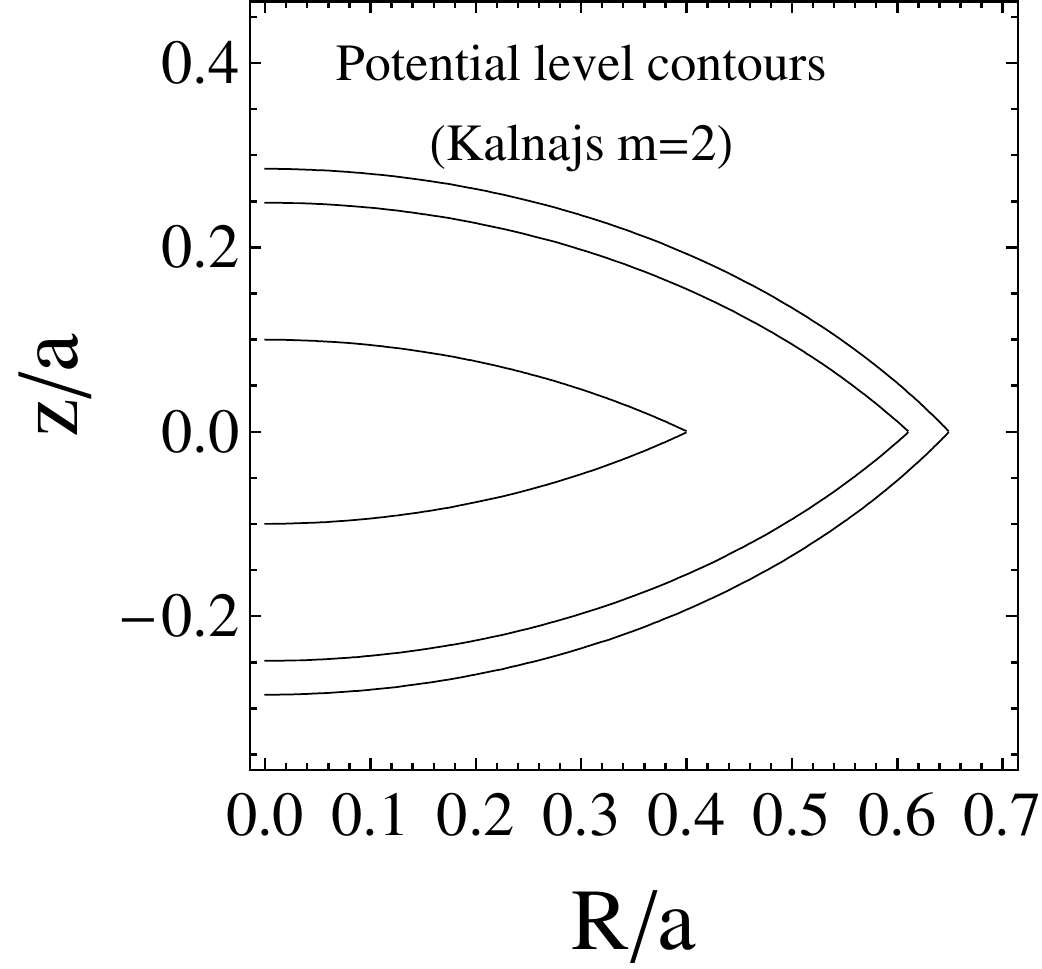,width=0.33\columnwidth ,angle=0}
\epsfig{figure=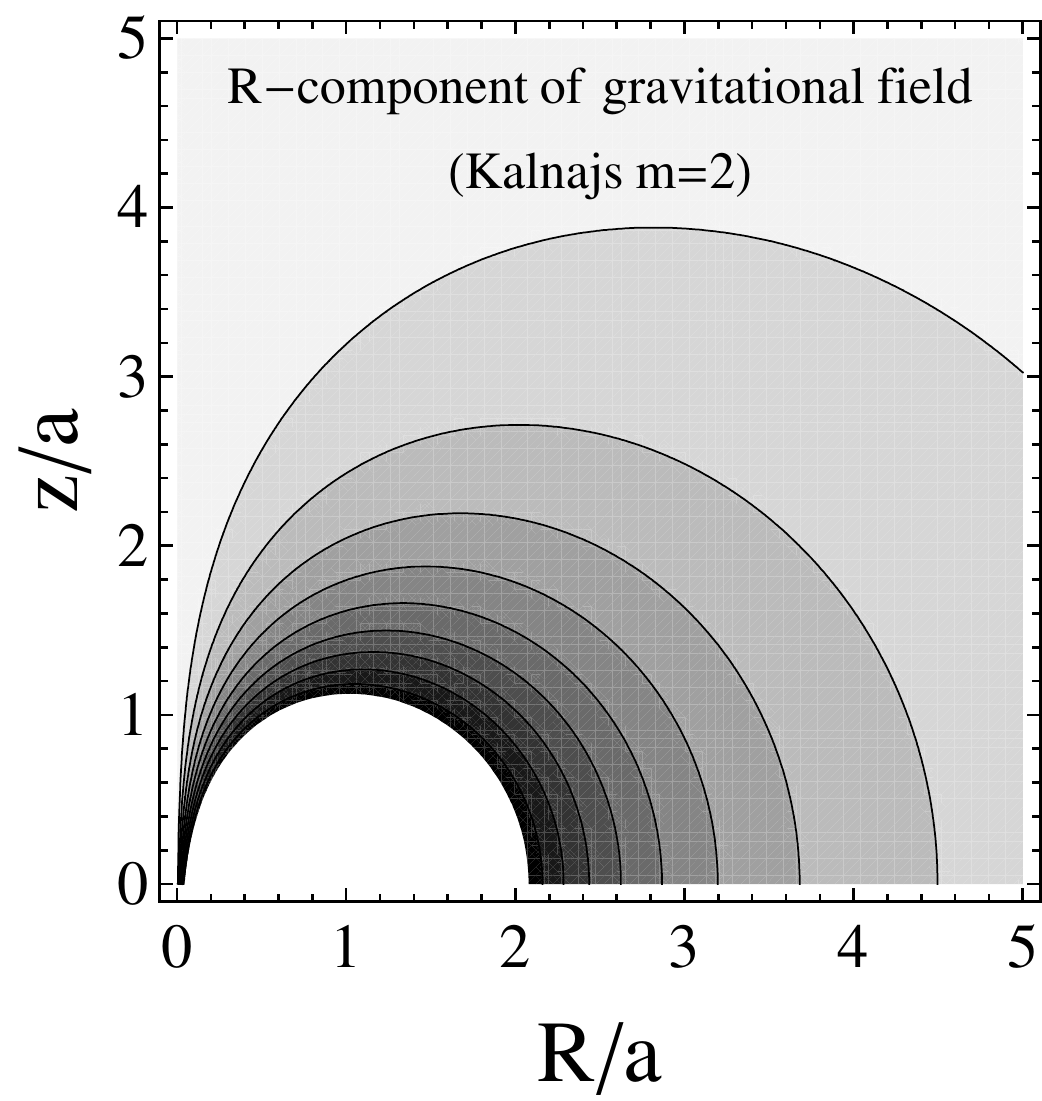,width=0.31\columnwidth ,angle=0}\quad
\epsfig{figure=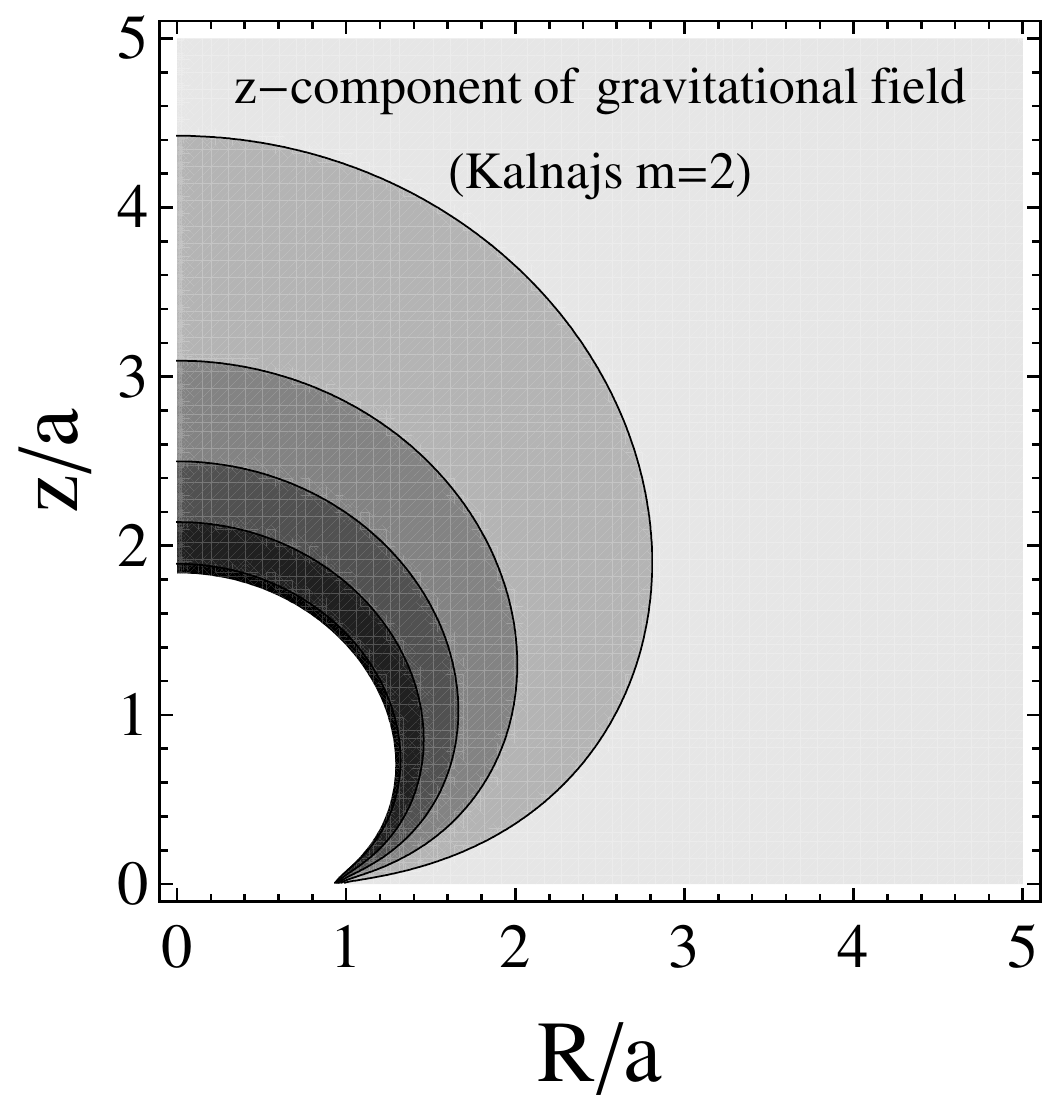,width=0.31\columnwidth ,angle=0}
\caption{
We show the potential and the modulus of $R$ and $z$ components of gravitational field 
(force per unit mass) corresponding to the two disks considered here: Kuzmin (top panels) and Kalnajs $m=2$
(bottom panels). Darker regions show higher values of the components, except for white central regions representing its maximum values.
}
\label{fig:Pot-force}
\end{figure}

We present the results of numerically calculated orbits in
three cases: (i) the Kuzmin disk, (ii) the second member of the family of generalized Kalnajs disks (in figure \ref{fig:Pot-force} we illustrate
the potential and gravitational field for these models)
 and (iii) the Kuzmin disk surrounded by a Plummer halo.
By introducing the halo, one might see how the surrounding structure around the disk interferes in the prediction of (\ref{eq:ZZ'sigma}),
which depends solely on the disk's surface density.

\subsection{Orbits in Kuzmin disk}\label{sec:numericalKuz}

\begin{figure}
\epsfig{figure=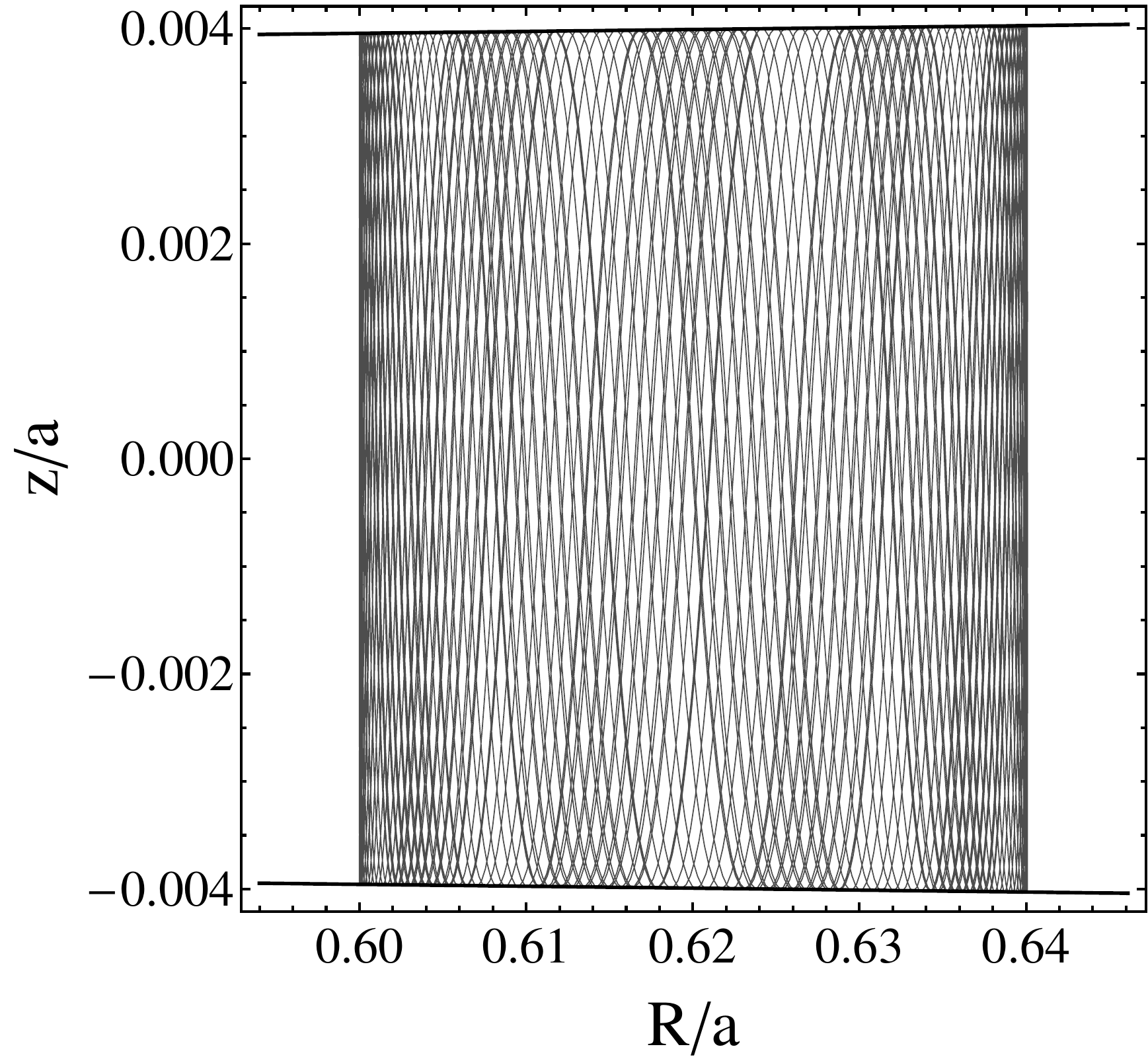,width=0.48\columnwidth ,angle=0}\quad
\epsfig{figure=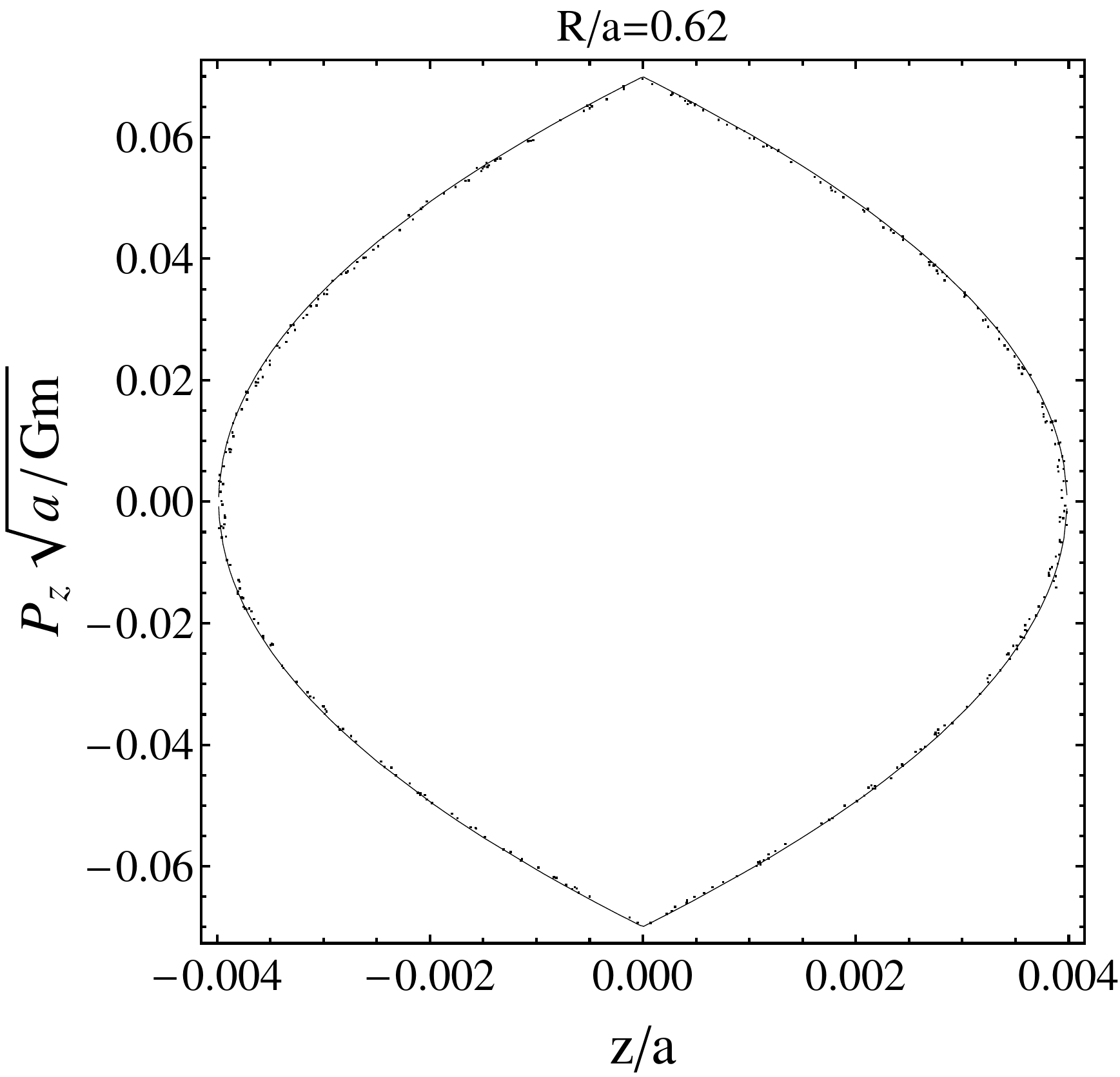,width=0.48\columnwidth ,angle=0}\\ \\
\epsfig{figure=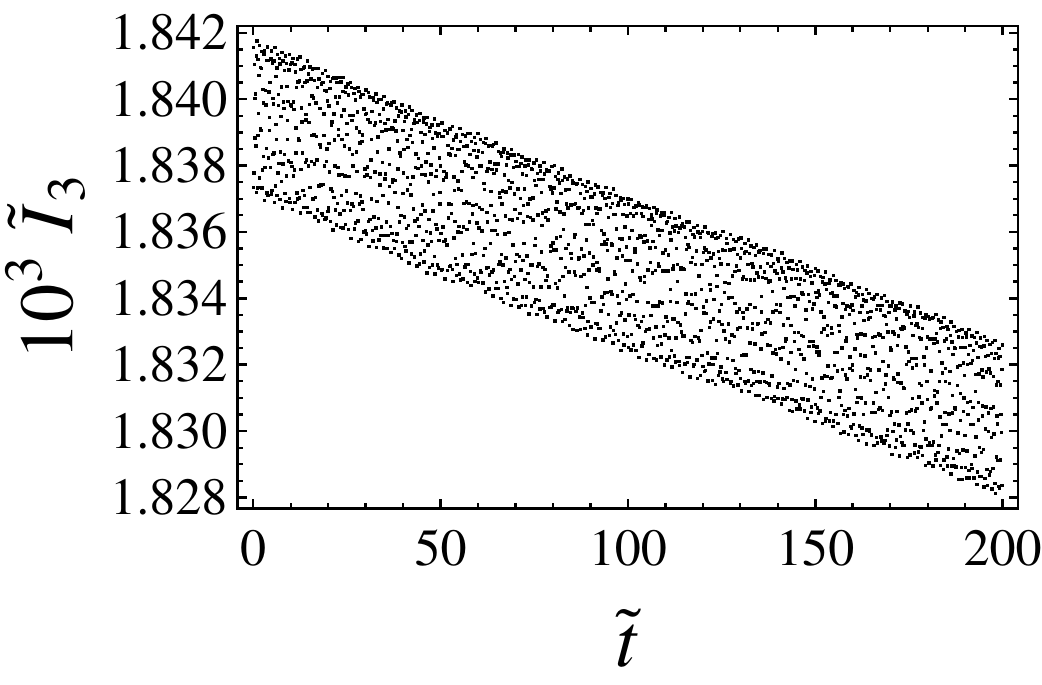,width=0.47\columnwidth ,angle=0}\quad
\epsfig{figure=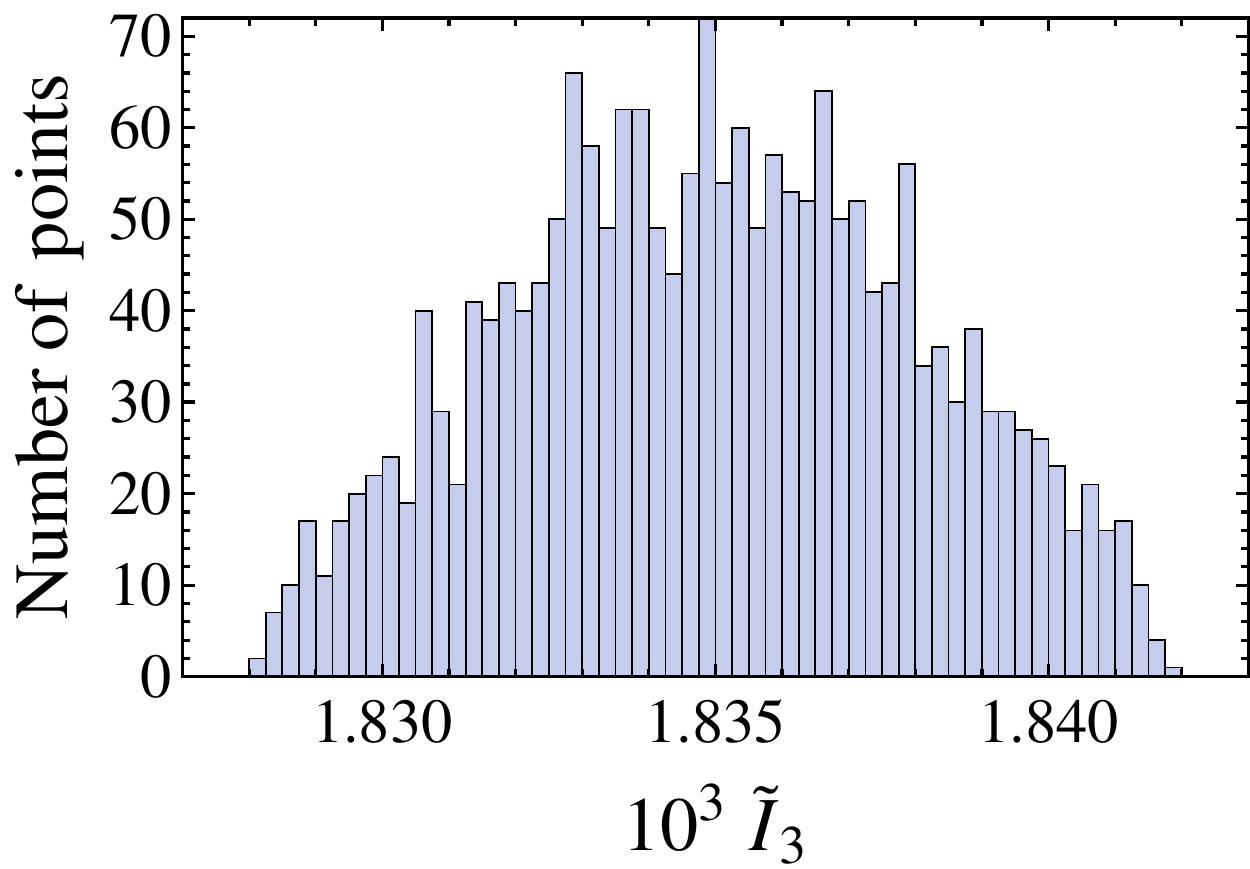,width=0.48\columnwidth ,angle=0}
\caption{Top, left: Orbit in Kuzmin's potential shown in the meridional plane.
It has initial conditions $R_{o}/a=0.6$, $P_{R}=0$ and $z_{o}/a=10^{-15}$.
The values of angular momentum and energy are $\ell/\sqrt{aGm}=0.3$, $aE/Gm=-0.73$, giving $\Delta E=3.9\times 10^{-3}$.
The ratio between the vertical and radial average periods is $T_z/T_R\approx0.05$.
It was calculated from the peaks in the time series of $z(t)$ and $R(t)$.
The predicted envelopes from Eq.~(\ref{eq:ZZ'sigma}) are shown by black lines.
In our choice, the predicted and numerically calculated envelopes have the same value
at the upper point of the zero-velocity curve. The prediction agrees with the envelopes of the numerically integrated orbits.
Top, right: Poincar\'e section in the surface of constant energy and angular momentum given by the orbit's parameters, with $R/a=0.62$.
The consequents of the orbits are calculated for both $P_R>0$ and $P_R<0$ and are
given by black dots, whereas the prediction from the average value of $\tilde{I}_{3,\rm mean}=1.835\times 10^{-3}$ is given by the solid black line.
We use the dimensionless quantities $\tilde I_3=(aM)^{-1/3} I_3$ and $\tilde t = (GM/a^3)^{1/2}t$.
We see that Eq.~(\ref{eq:I3integral}) is a good description for the approximate third integral in low-amplitude orbits.
Bottom, left: Time series of $\tilde I_3(t)$, given by Eq.~(\ref{eq:I3integral}).
It shows small spread along time evolution.
Bottom, right: Histogram for $\tilde I_3(t)$.}
\label{fig:Kuz1pp}
\end{figure}

\begin{figure}
\epsfig{figure=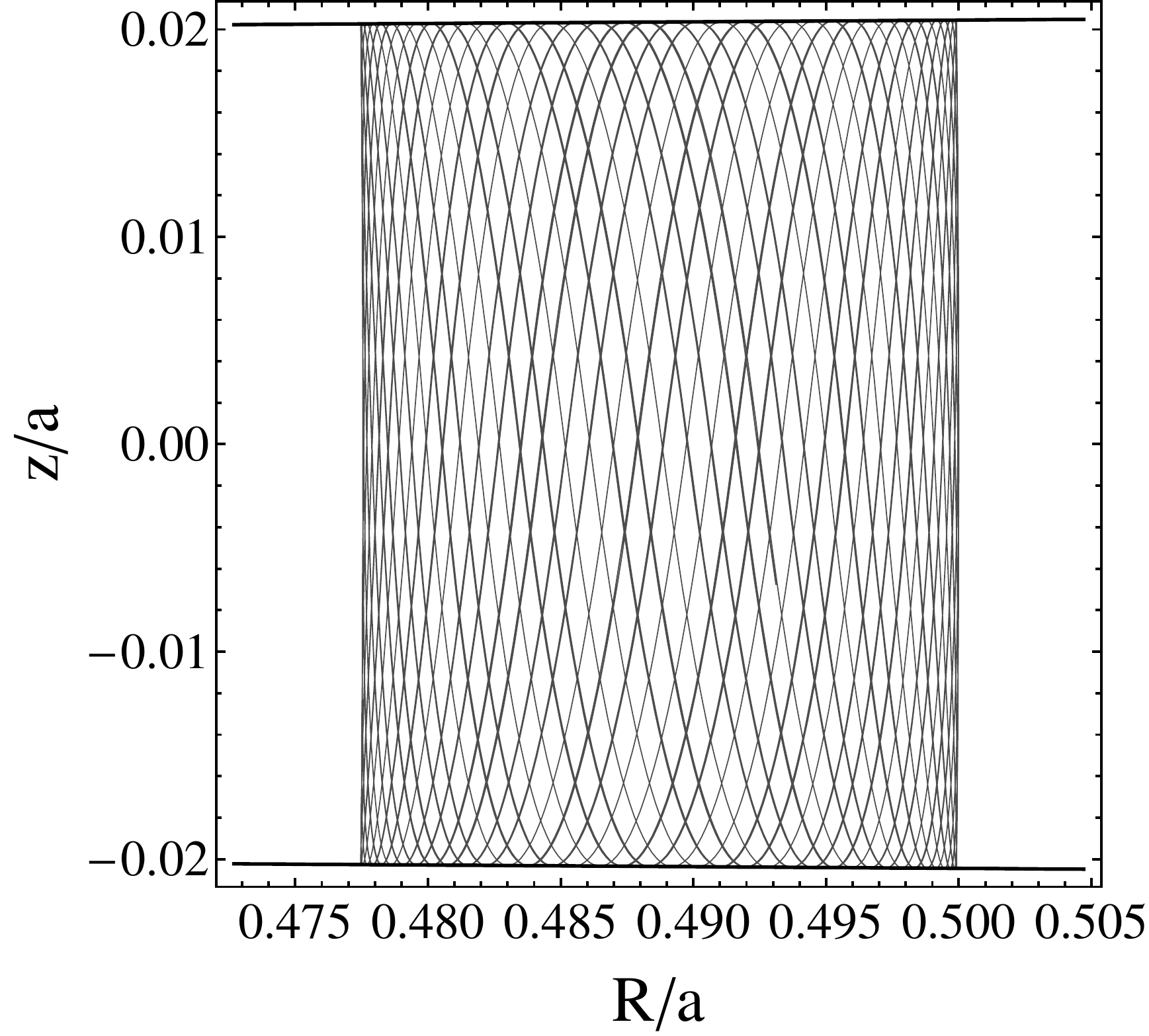,width=0.48\columnwidth ,angle=0}\quad
\epsfig{figure=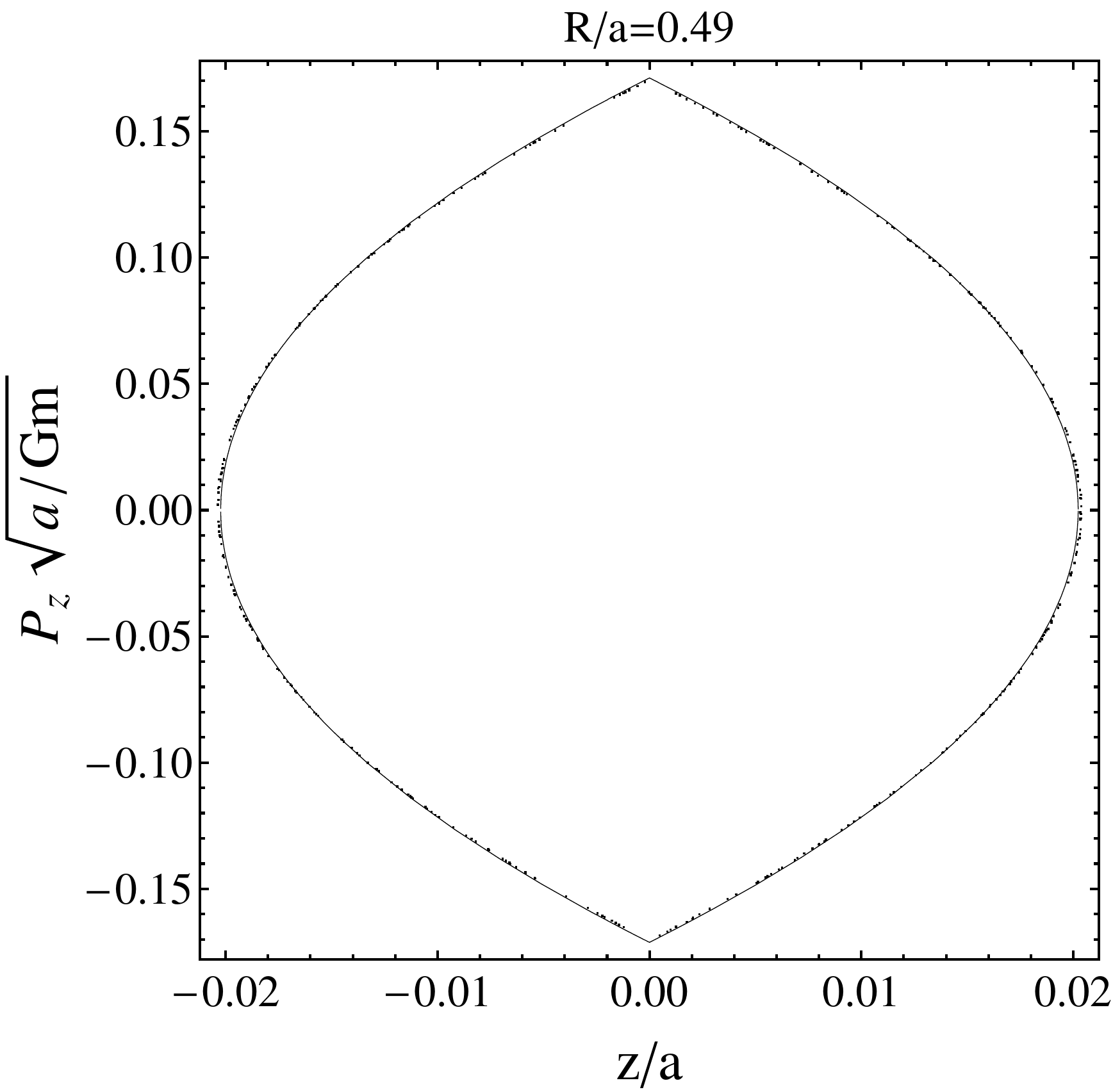,width=0.48\columnwidth ,angle=0}\\ \\
\epsfig{figure=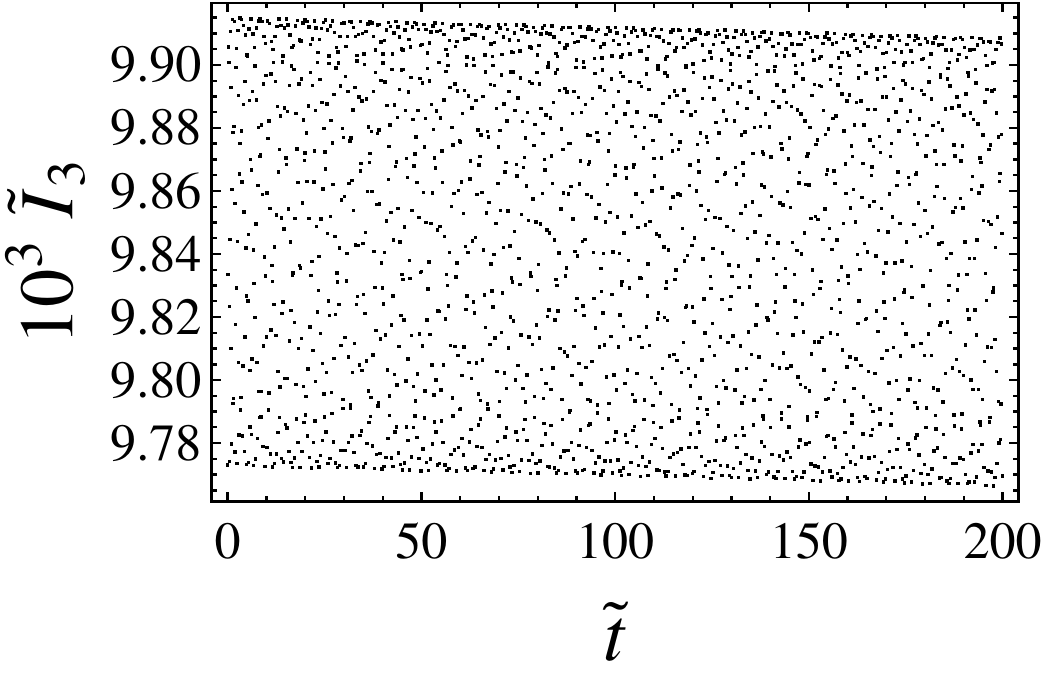,width=0.47\columnwidth ,angle=0}\quad
\epsfig{figure=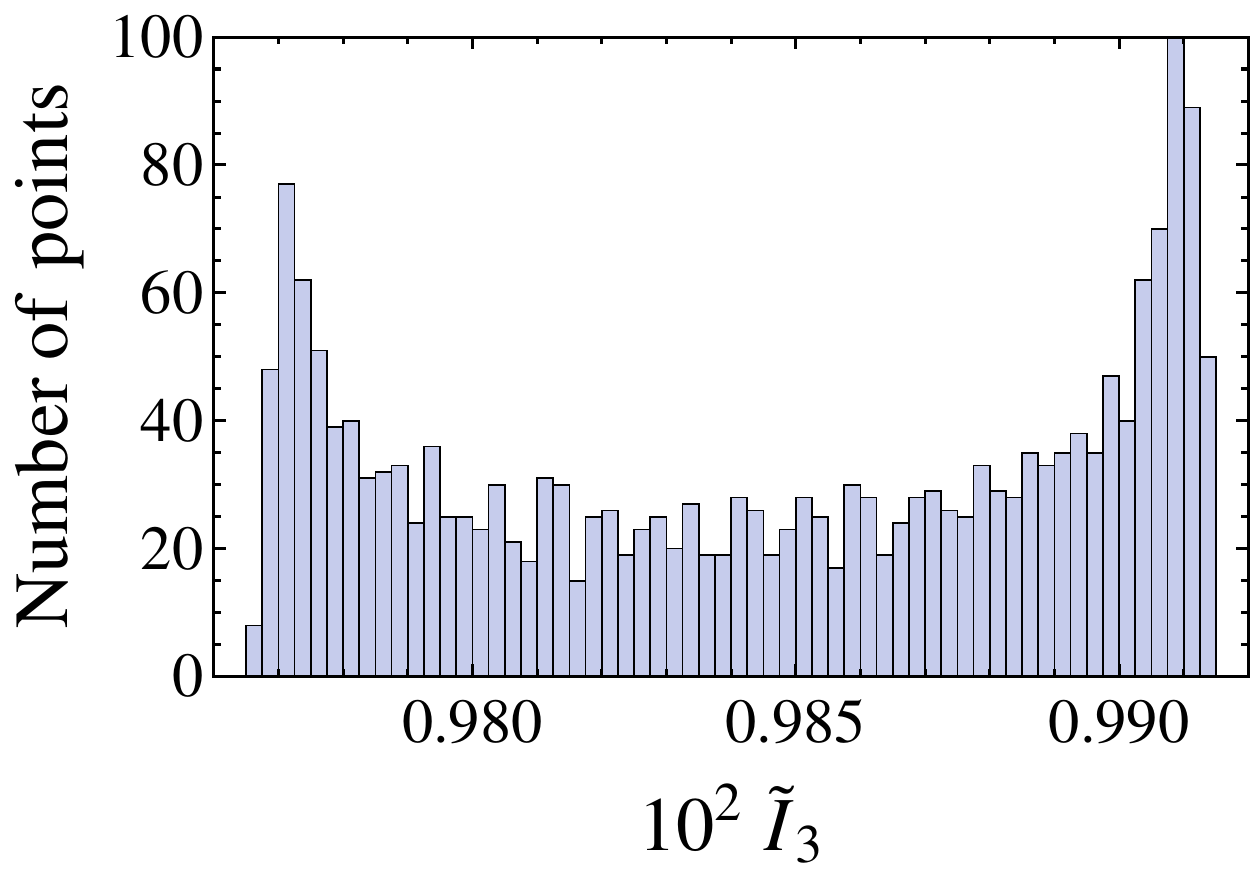,width=0.48\columnwidth ,angle=0}
\caption{Orbit in Kuzmin's potential with $R_{o}/a=0.5$, $P_{R}=0$ and $z_{o}/a=10^{-9}$.
The values of angular momentum and energy are $\ell/\sqrt{aGm}=0.2$, $aE/Gm=-0.8$, giving $\Delta E=1.8\times 10^{-2}$.
We have $T_z/T_R\approx0.15$. The Figure's style is the same as in Fig.~\ref{fig:Kuz1pp}.
The Poincar\'e section is calculated for $R/a=0.49$.
We see that the prediction of Eq.~(\ref{eq:I3integral}) is still a good
approximation for this value of vertical amplitude and $\Delta E$.
The spread in $I_3(t)$ remains small and the consequents in the Poincar\'e section
lie practically on the curve predicted by  $\tilde{I}_{3,\rm mean}=9.845\times 10^{-3}$.}
\label{fig:Kuz1peq}
\end{figure}

\begin{figure}
\epsfig{figure=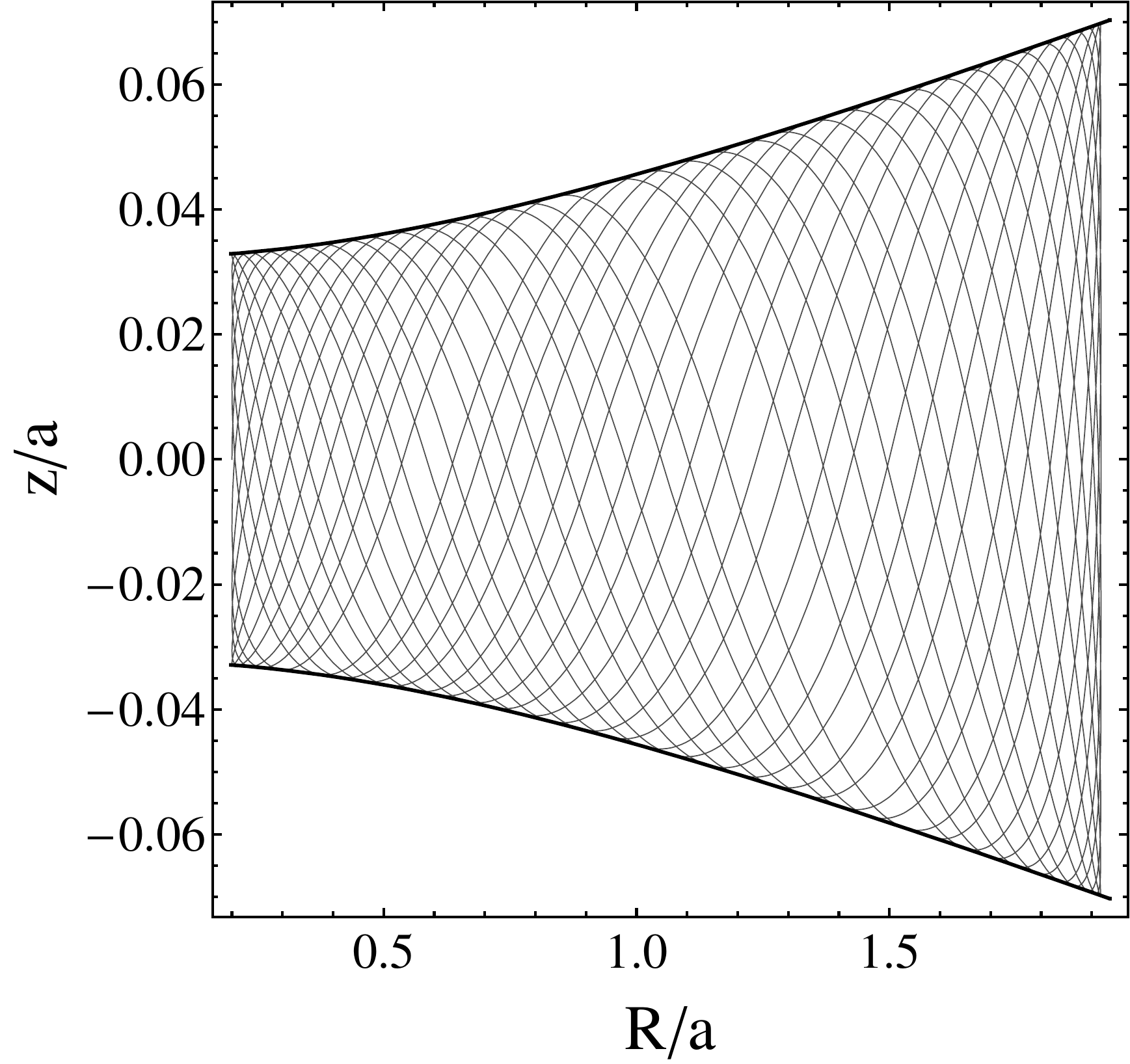,width=0.48\columnwidth ,angle=0}\quad
\epsfig{figure=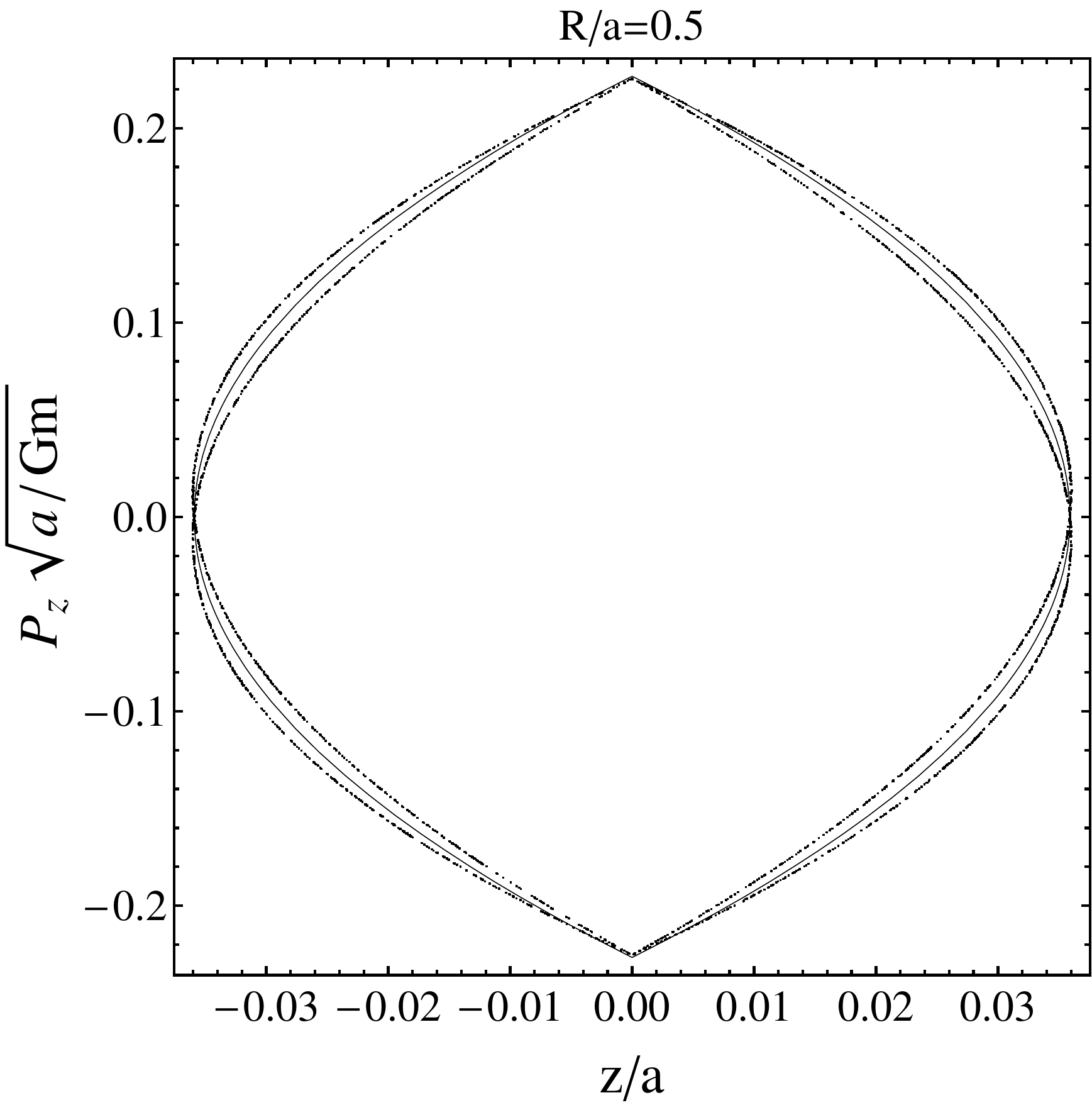,width=0.48\columnwidth ,angle=0}\\ \\
\epsfig{figure=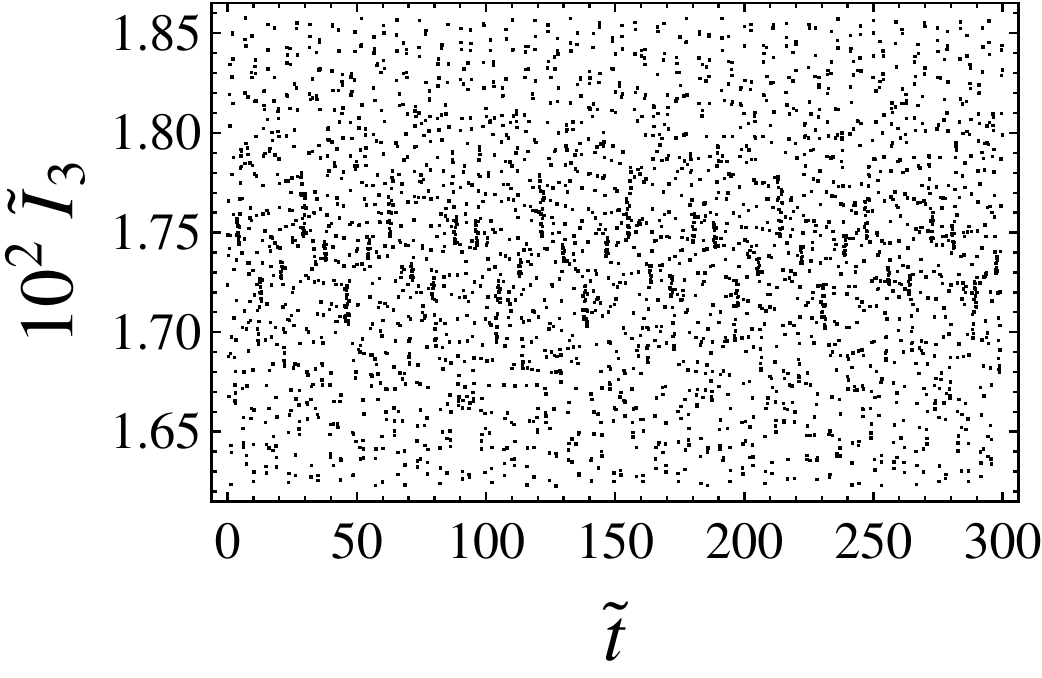,width=0.47\columnwidth ,angle=0}\quad
\epsfig{figure=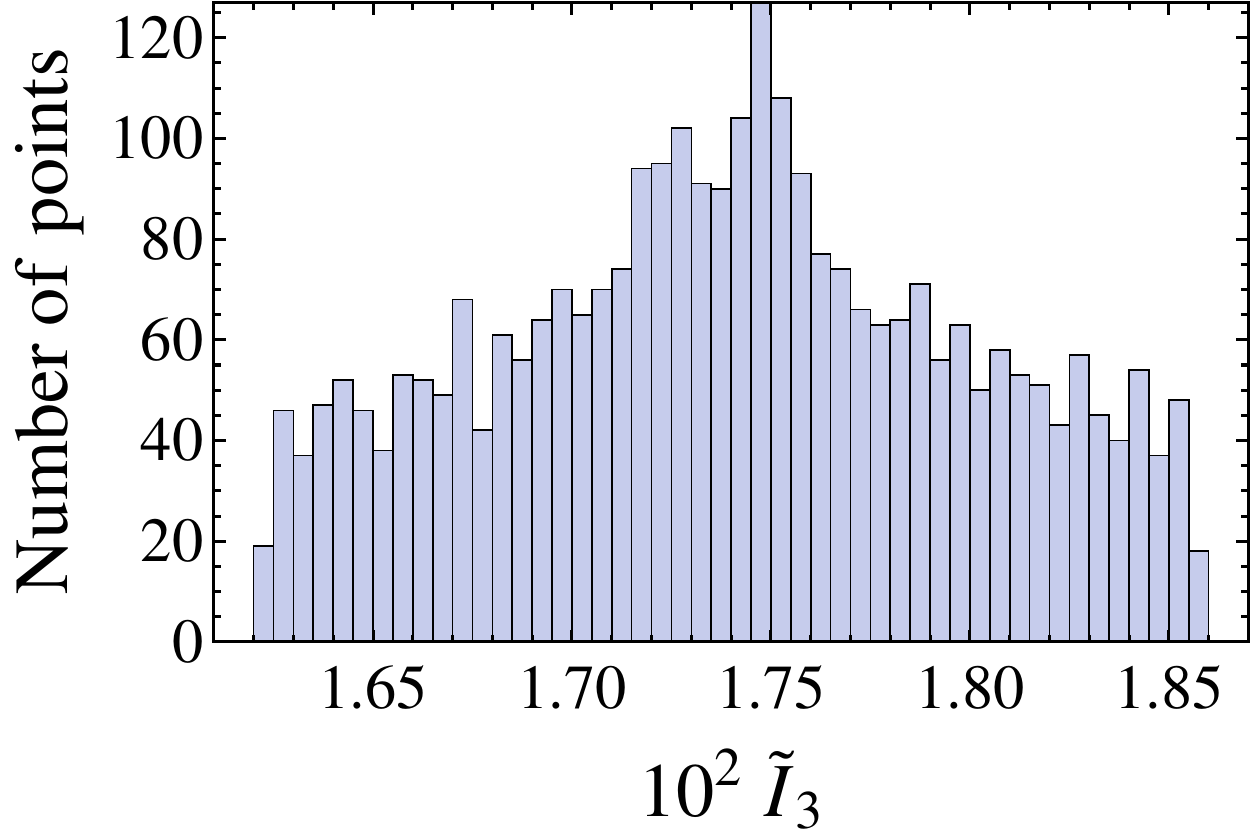,width=0.48\columnwidth ,angle=0}
\caption{Orbit in Kuzmin's potential with $R_{o}/a=0.2$, $P_{R}=0$ and $z_{o}/a=10^{-10}$.
The values of angular momentum and energy are $\ell/\sqrt{aGm}=0.2$, $aE/Gm=-0.45$, giving $\Delta E=0.4477$.
We have $T_z/T_R=0.29$. The Figure's style is the same as in Fig.~\ref{fig:Kuz1pp}.
We see that the spread in the approximate third integral is larger than in
Figs.~\ref{fig:Kuz1pp} and \ref{fig:Kuz1peq}, because the orbits are deviating from
quasi-circular motion. The Poincar\'e section is calculated for $R/a=0.5$ and $\tilde{I}_{3,\rm mean}=1.74\times 10^{-2}$.
The consequents of the orbit are different for $P_R>0$ and $P_R<0$, since the orbit has a larger vertical
amplitude (see e.g. \citealp{binneyMcmillan2011MNRAS}).}
\label{fig:Kuz1a}
\end{figure}	

\begin{figure}
\epsfig{figure=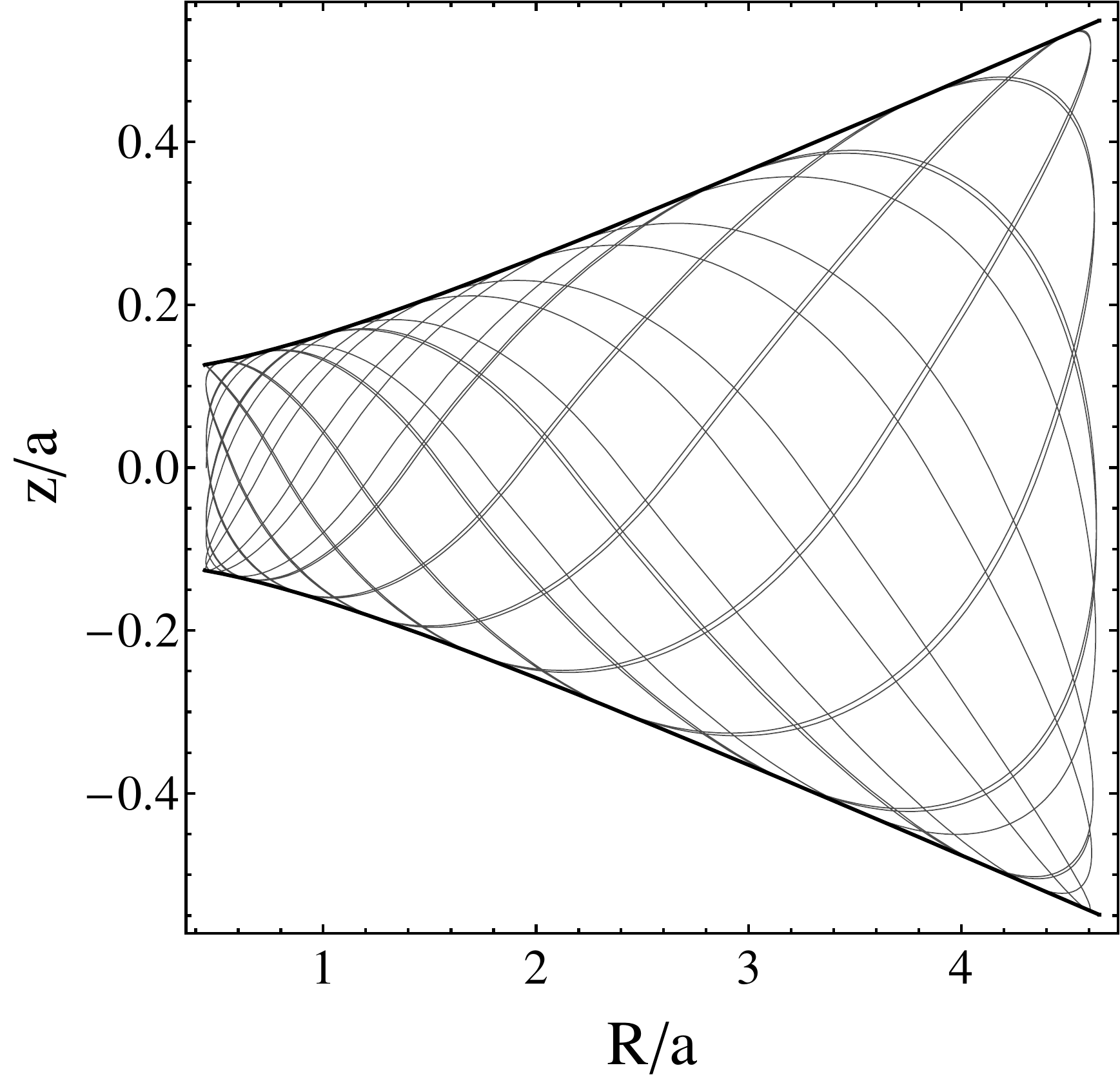,width=0.48\columnwidth ,angle=0}\quad
\epsfig{figure=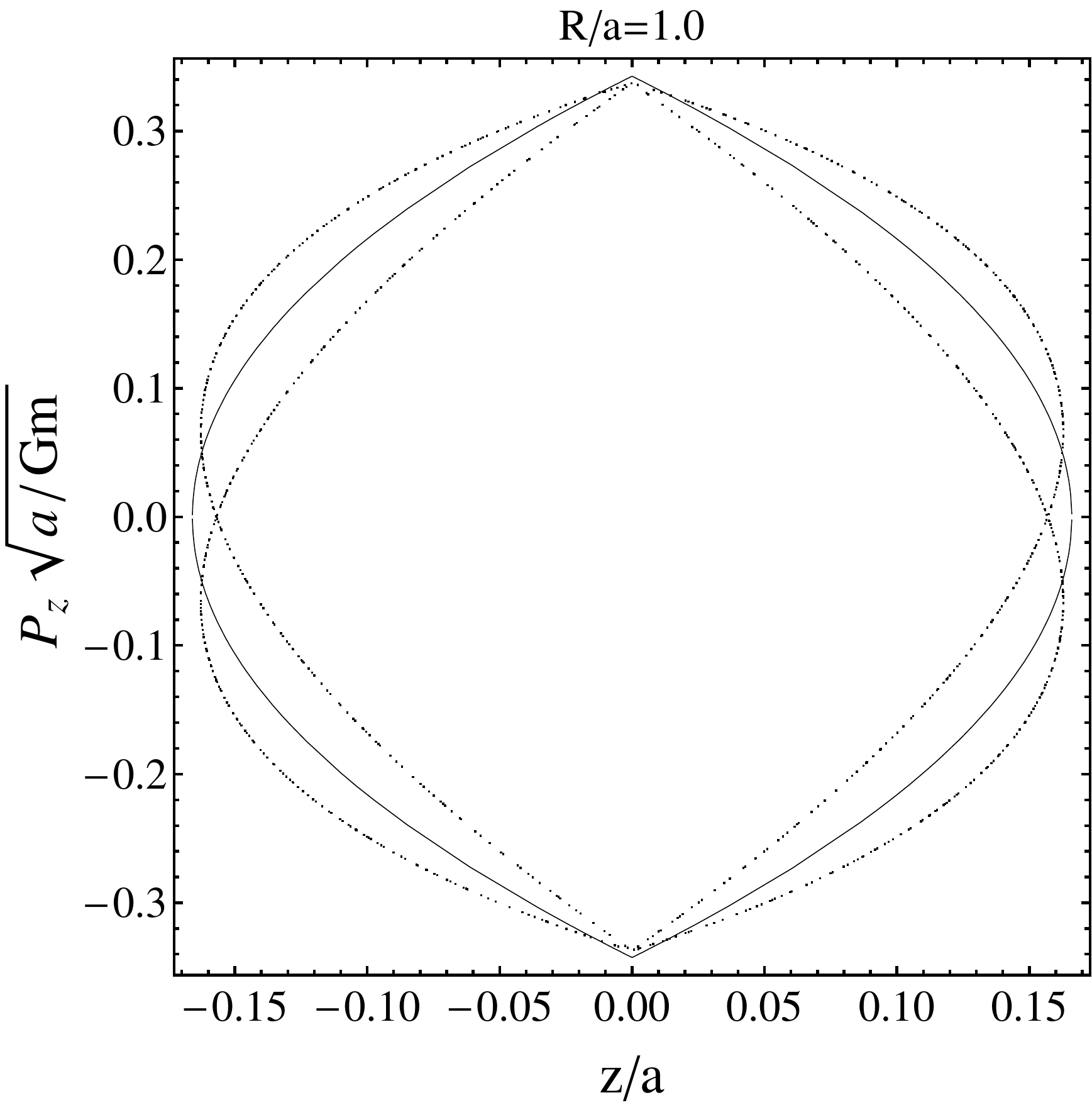,width=0.48\columnwidth ,angle=0}\\
\epsfig{figure=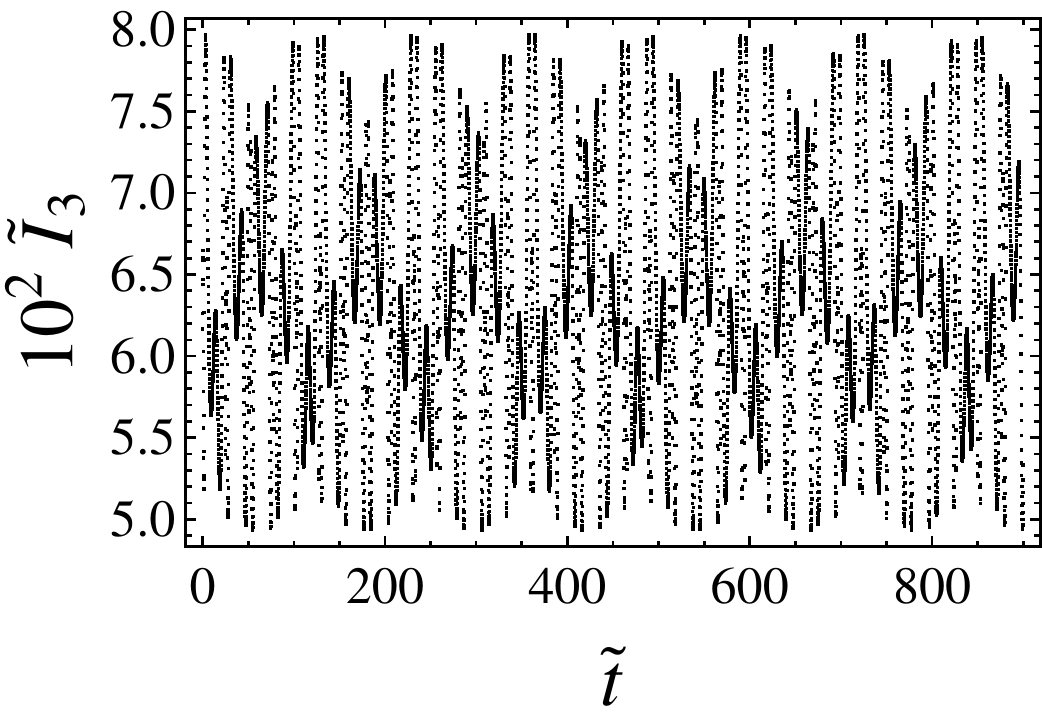,width=0.47\columnwidth ,angle=0}\quad
\epsfig{figure=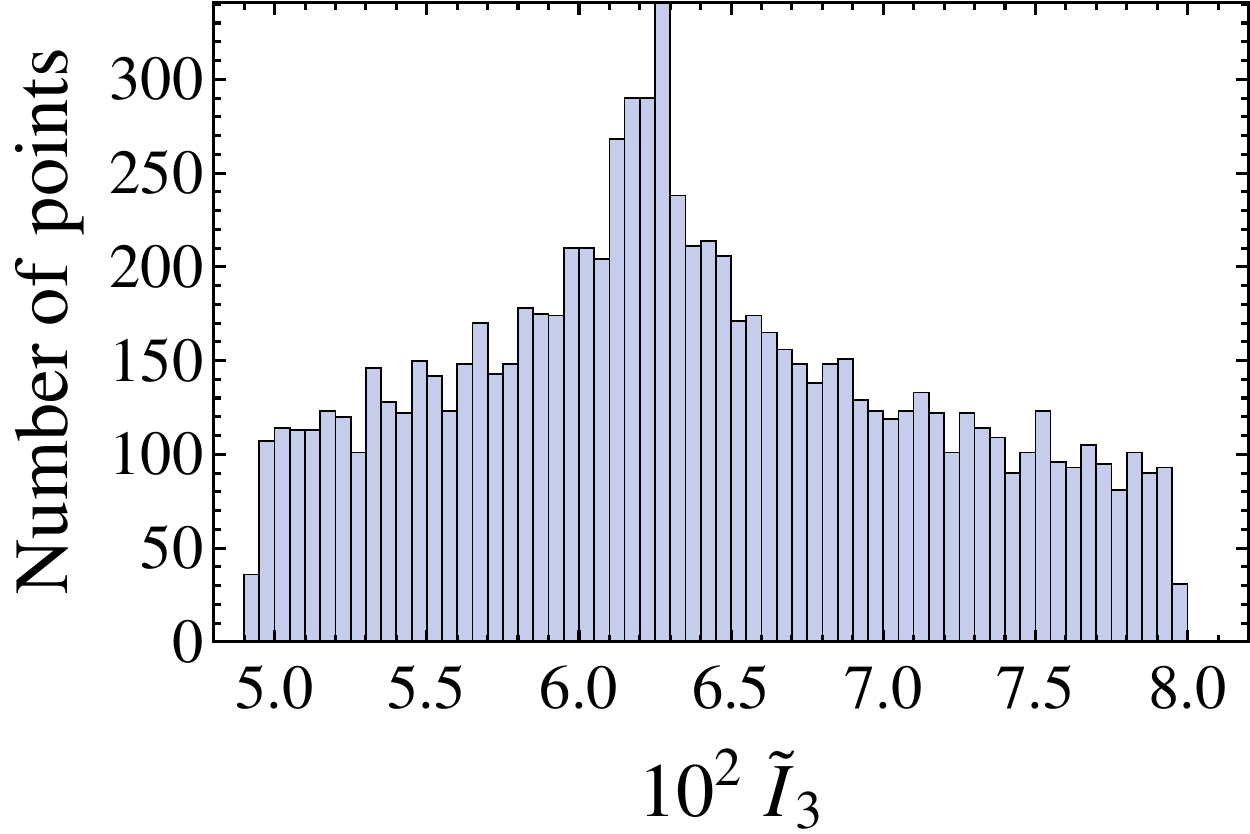,width=0.48\columnwidth ,angle=0}
\caption{Orbit in Kuzmin's potential with $R_{o}/a=0.45$, $P_{R}=0$ and $z_{o}/a=10^{-10}$.
The values of angular momentum and energy are $\ell/\sqrt{aGm}=0.5$, $aE/Gm=-0.2$, giving $\Delta E=0.6606$.
We have $T_z/T_R\approx0.54$. The Figure's style is the same as in Fig.~\ref{fig:Kuz1pp}.
The Poincar\'e section is calculated for $R/a=1.0$. We see that the spread in the approximate third integral
is larger than in Fig.~\ref{fig:Kuz1a}. The difference between the consequents of the orbit in the Poincar\'e
section and the prediction from $\tilde{I}_{3,\rm mean}=6.36\times 10^{-2}$ also increases comparing to Fig.~\ref{fig:Kuz1a}.
This is due, among other factors, to the deviations from quasi-circular motion and the fact that $T_z/T_R$ approaches 1.}
\label{fig:Kuz1big}
\end{figure}

\begin{figure}
\epsfig{figure=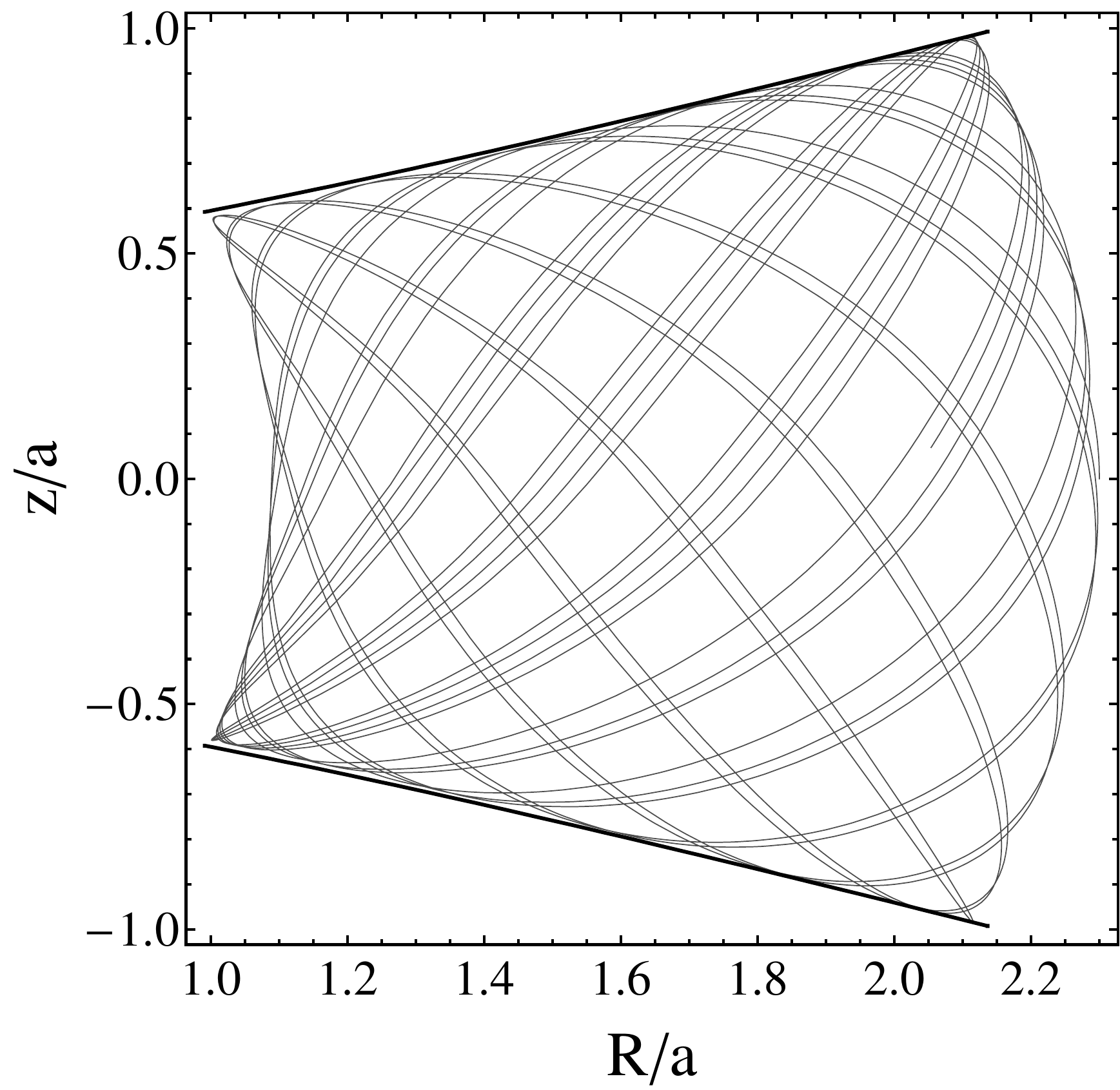,width=0.48\columnwidth ,angle=0}\quad
\epsfig{figure=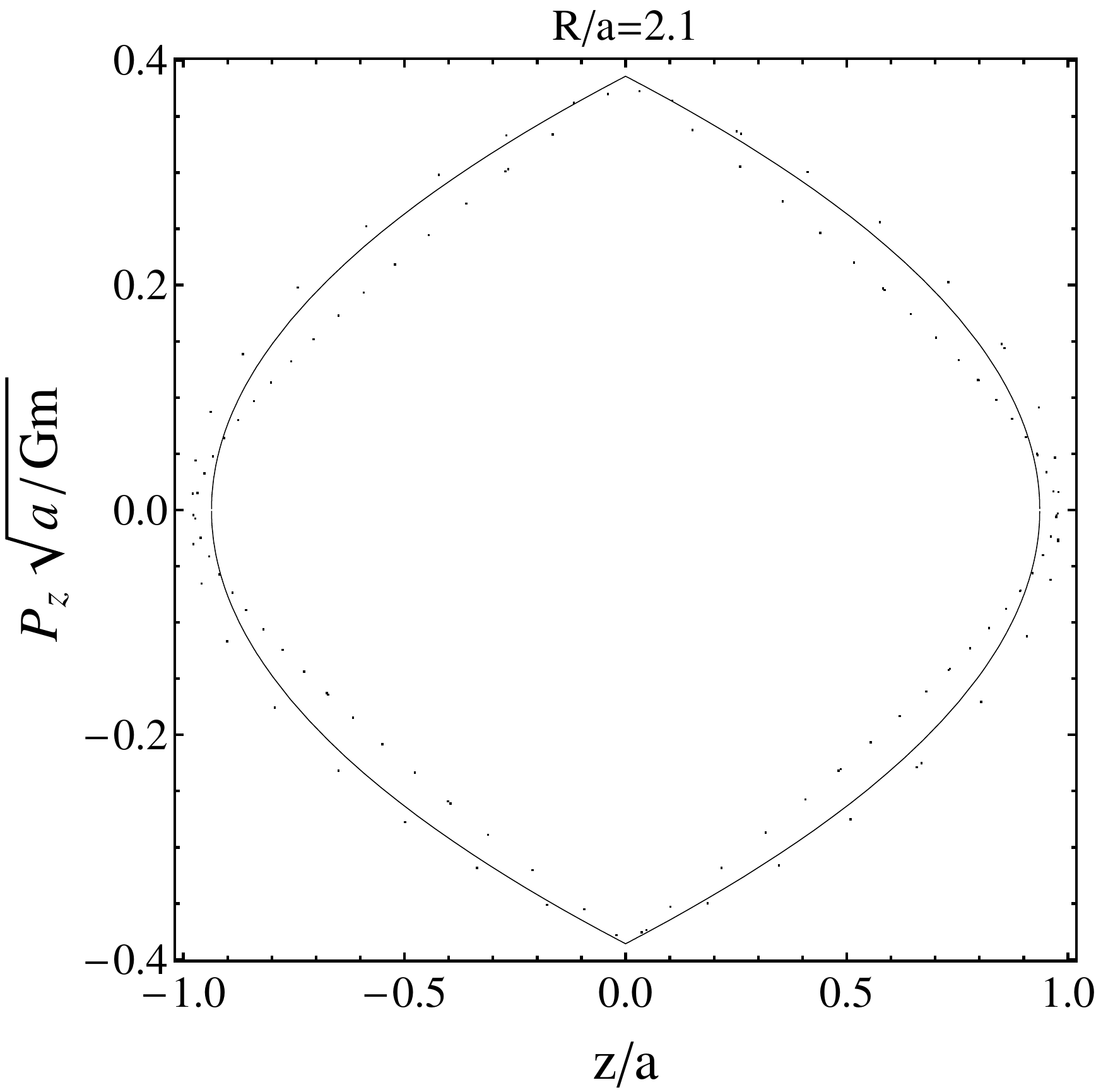,width=0.48\columnwidth ,angle=0}\\ \\
\epsfig{figure=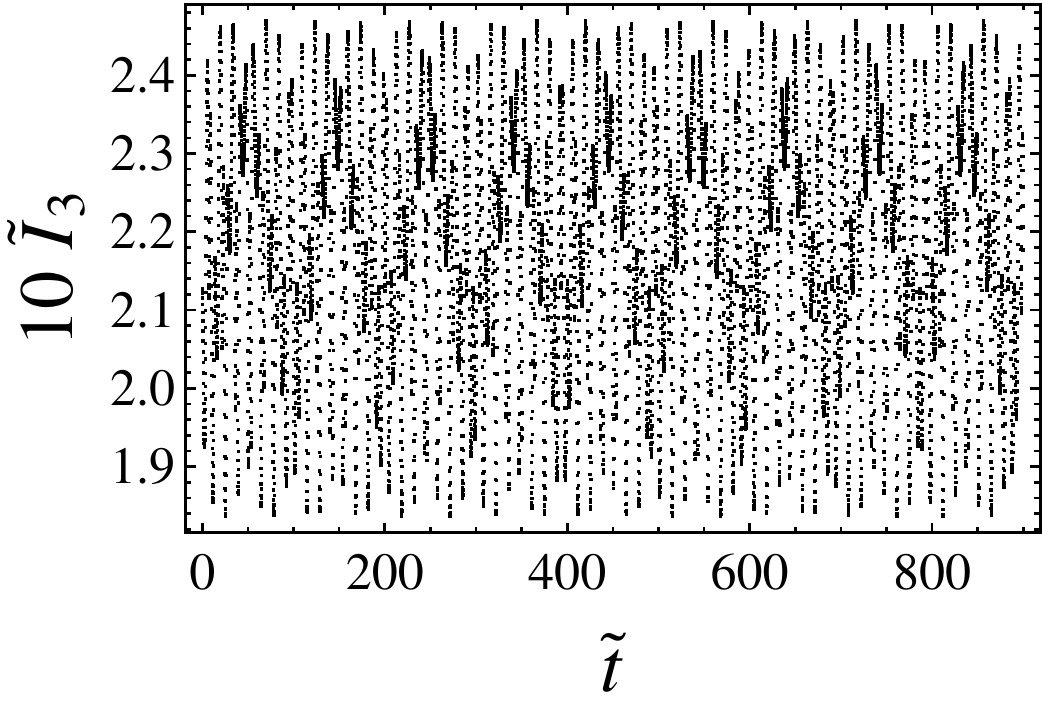,width=0.47\columnwidth ,angle=0}\quad
\epsfig{figure=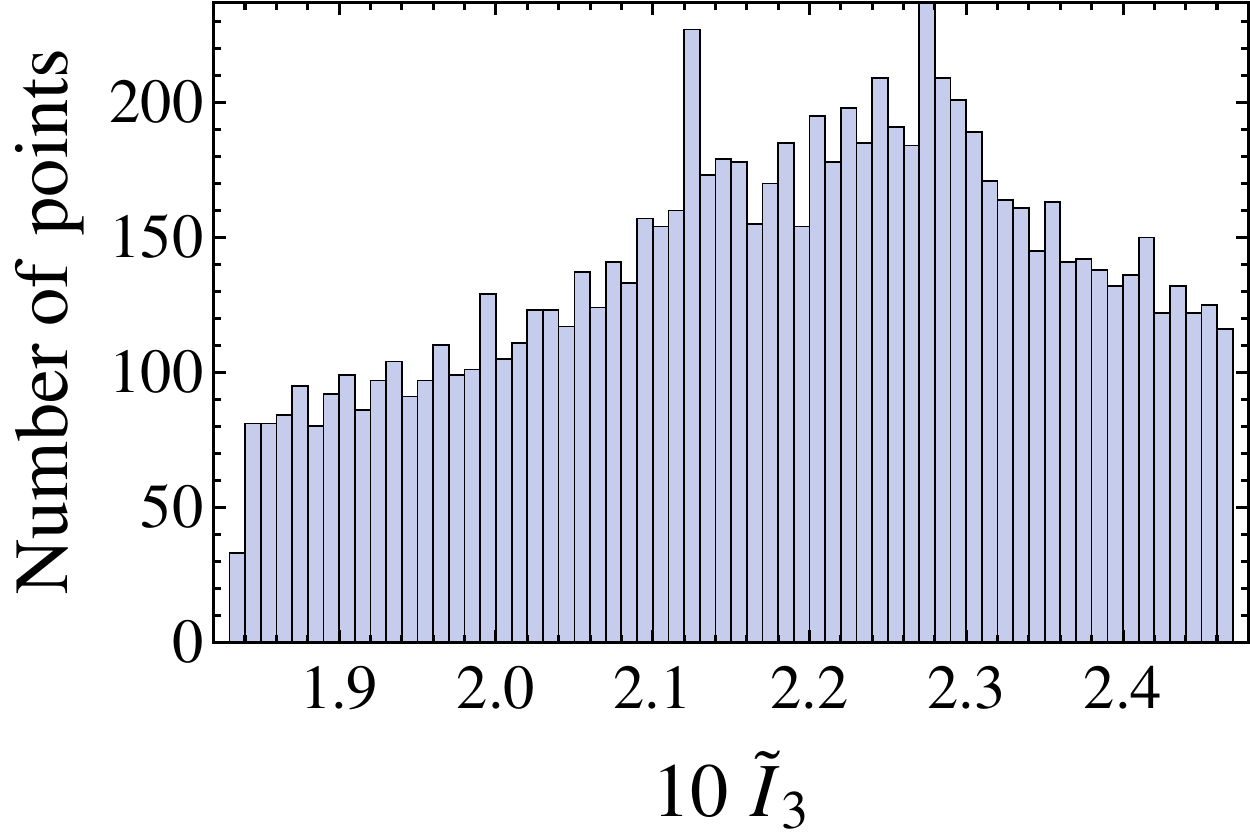,width=0.48\columnwidth ,angle=0}
\caption{Orbit in Kuzmin's potential with $R_{o}/a=2.3$, $P_{R}=0$ and $z_{o}/a=10^{-10}$.
The values of angular momentum and energy are $\ell/\sqrt{aGm}=0.7$, $aE/Gm=-0.29$, giving $\Delta E=0.384$.
We have $T_z/T_R\approx0.91$. The Figure's style is the same as in Fig.~\ref{fig:Kuz1pp}.
The Poincar\'e section is calculated for $R/a=2.1$ and $\tilde{I}_{3,\rm mean}=0.2181$.}
\label{fig:Kuz1c}
\end{figure}

The potential-density pair for the Kuzmin disk,  $(\Phi_k,\Sigma_k)$, is given by \citep{binney-tremaine:GD}
  \begin{equation}
   \Phi_k = - \frac{Gm}{\sqrt{R^2 + (|z|+a)^2}},\qquad
   \Sigma_k = \frac{m}{2\pi a^{2}}\left(1+\frac{R^2}{ a^2} \right)^{-3/2},\label{PotDensKuz}
  \end{equation}
where $m$ is the total mass of the disk and $a$ is a positive parameter representing the radial length scale.
Figure~\ref{fig:Pot-force} top shows the potential level contours for $\Phi_k$ and the $R$- and $z$-components of the corresponding gravitational field.
We obtain, besides accurate predictions for low-amplitude orbits
(Figs. \ref{fig:Kuz1pp} and \ref{fig:Kuz1peq}), very energetic orbits
whose envelopes are well described by (\ref{eq:ZZ'sigma}) (Figs. \ref{fig:Kuz1a}, \ref{fig:Kuz1big} and \ref{fig:Kuz1c}).
These orbits can span a considerable radial range,
and their vertical amplitudes may be
comparable to the radial length scale of the disk, $Z/a\lesssim 1$.
All orbits of Figs. \ref{fig:Kuz1pp}--\ref{fig:Kuz1c} have $T_z/T_R<1$, calculated via the number of peaks in the time series for $z(t)$ and $R(t)$,
where $T_z \ (T_R)$ is the vertical (radial) average period of oscillation,
meaning that we are in the region where the adiabatic assumption is valid.
The precise reason for this unexpectedly wide range of validity for expressions
(\ref{eq:ZZ'sigma})--(\ref{eq:I3integral}) may be linked to the fact that the Kuzmin potential gives
rise to an integrable Hamiltonian \citep{hunter2005NYASA}.
We see next that, when considering a different razor-thin disk model, the range of validity of (\ref{eq:ZZ'sigma})--(\ref{eq:I3integral})
decreases abruptly.

\subsection{Orbits in Generalized Kalnajs disk $m=2$}\label{sec:numericalKalnajs}

\begin{figure}
\epsfig{figure=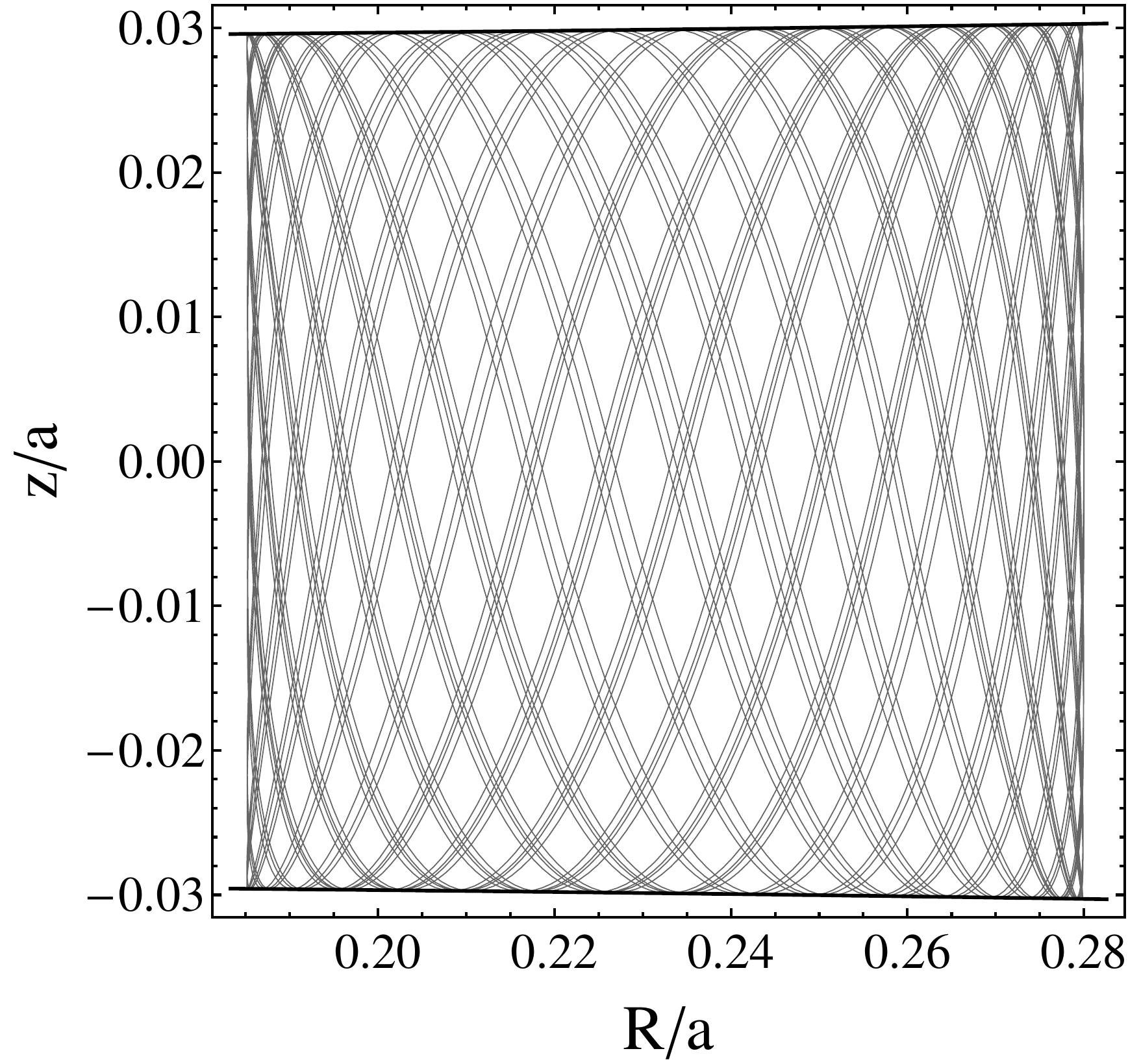,width=0.48\columnwidth ,angle=0}\quad
\epsfig{figure=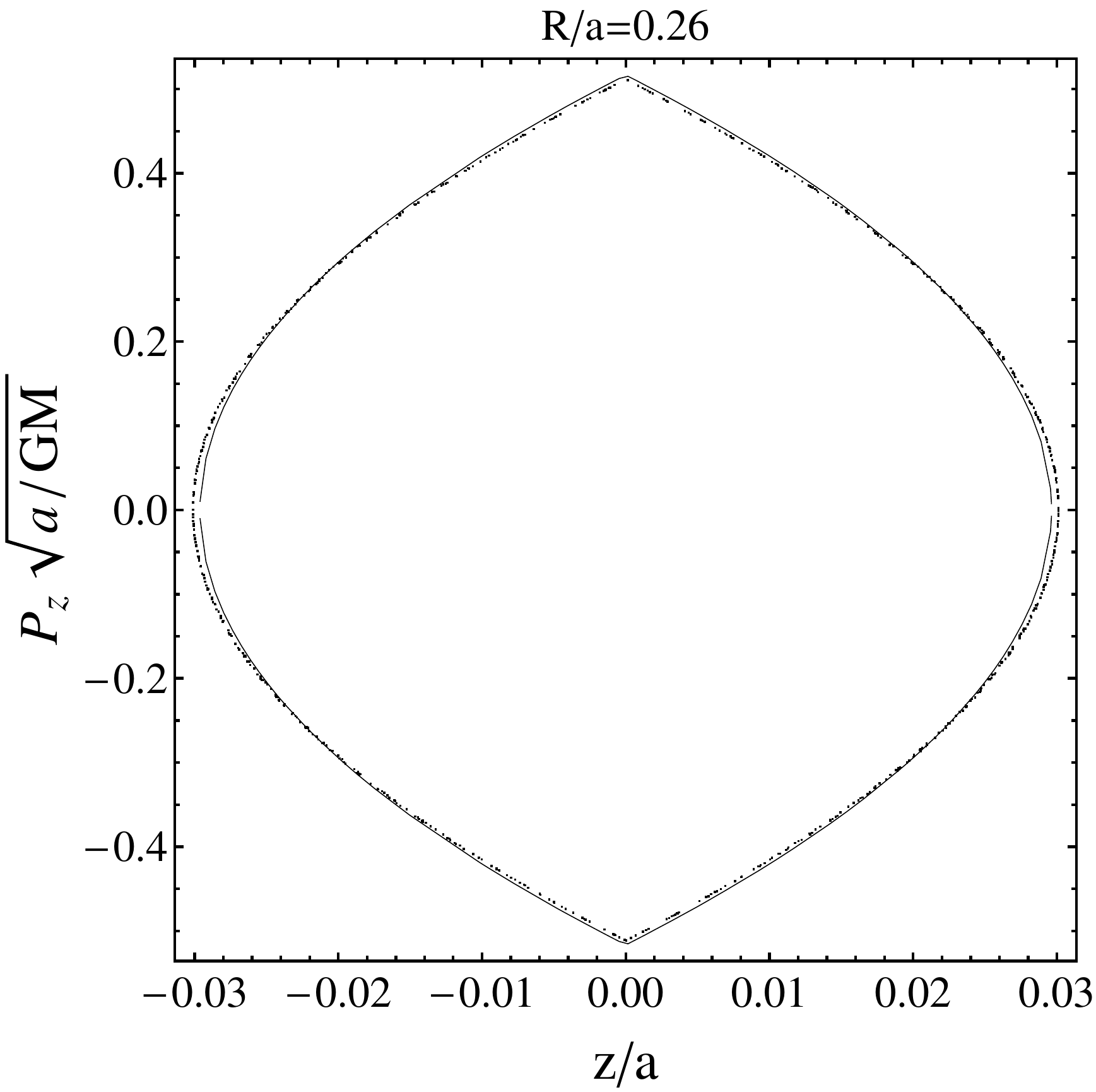,width=0.48\columnwidth ,angle=0}\\ \\
\epsfig{figure=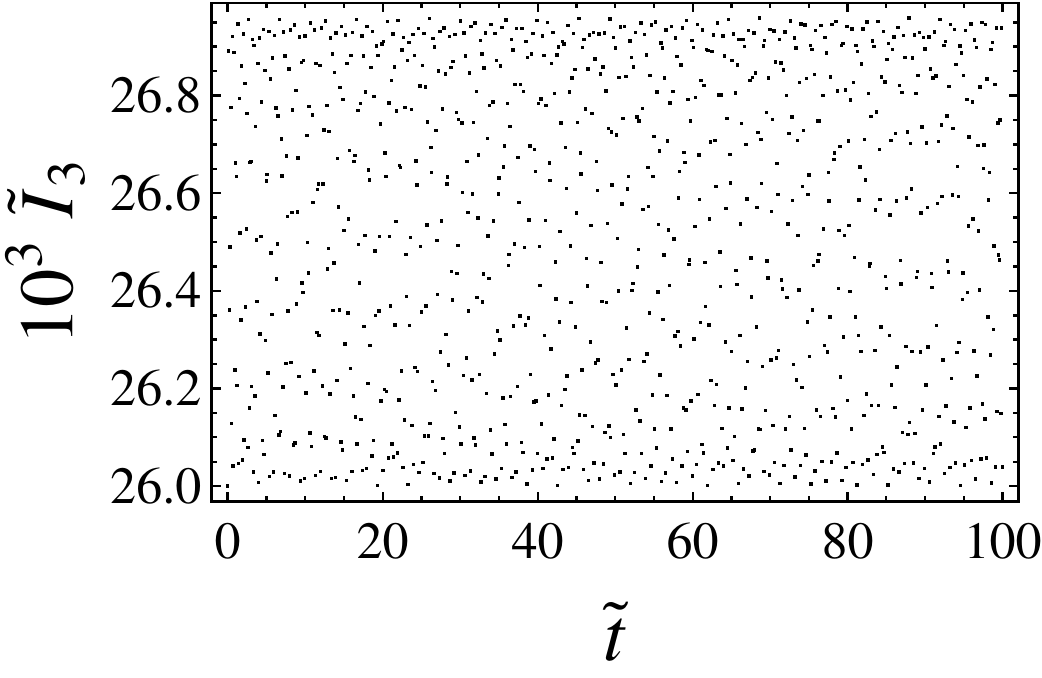,width=0.47\columnwidth ,angle=0}\quad
\epsfig{figure=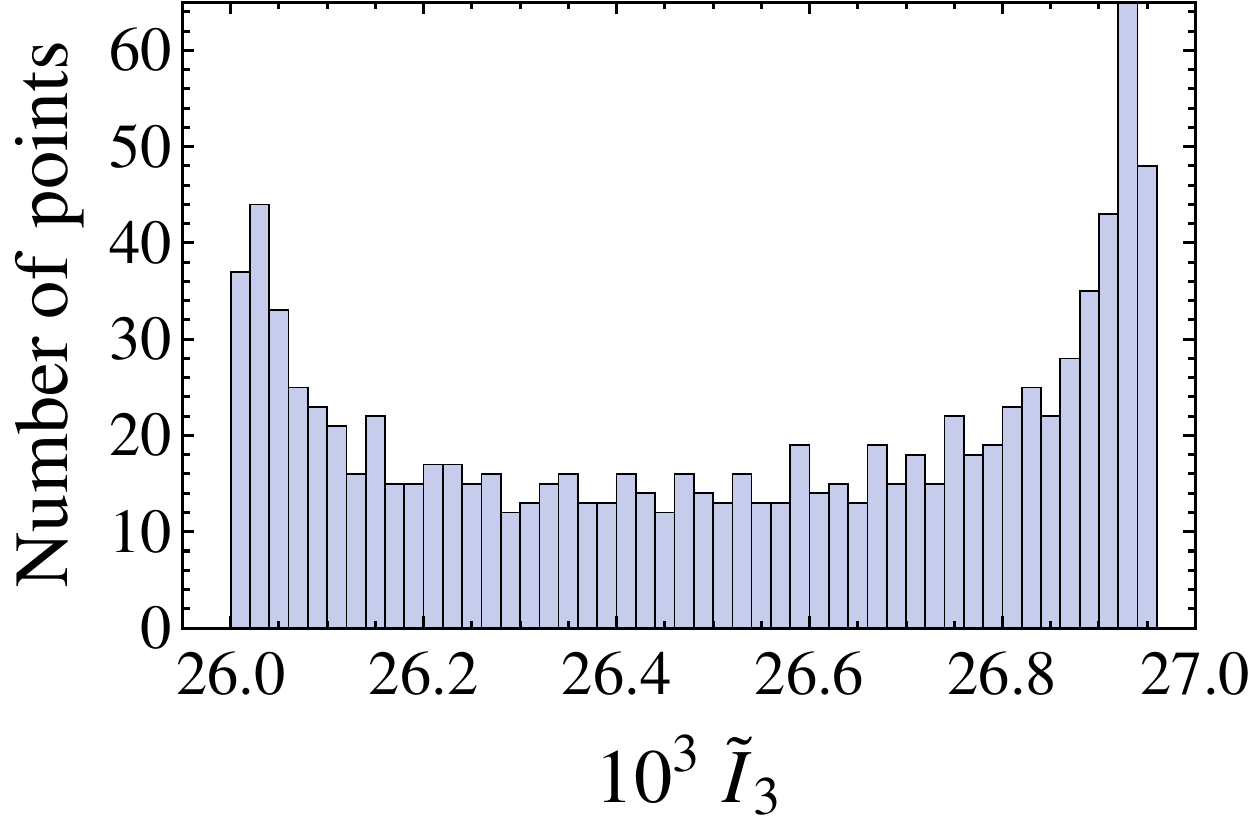,width=0.48\columnwidth ,angle=0}
\caption{Orbit in Kalnajs ($m=2$) potential with $R_{o}/a=0.28$, $P_{R}=0$ and $z_{o}/a=10^{-8}$.
The values of angular momentum and energy are $\ell/\sqrt{aGM}=0.12$, $aE/GM=-2.5$, giving $\Delta E=0.059$.
We have $T_z/T_R\approx0.33$. The Figure's style is the same as in Fig.~\ref{fig:Kuz1pp}.
The Poincar\'e section is calculated for $R/a=0.26$ and $\tilde{I}_{3,\rm mean}=0.0269$.}
\label{fig:K2a}
\end{figure}
\begin{figure}
\epsfig{figure=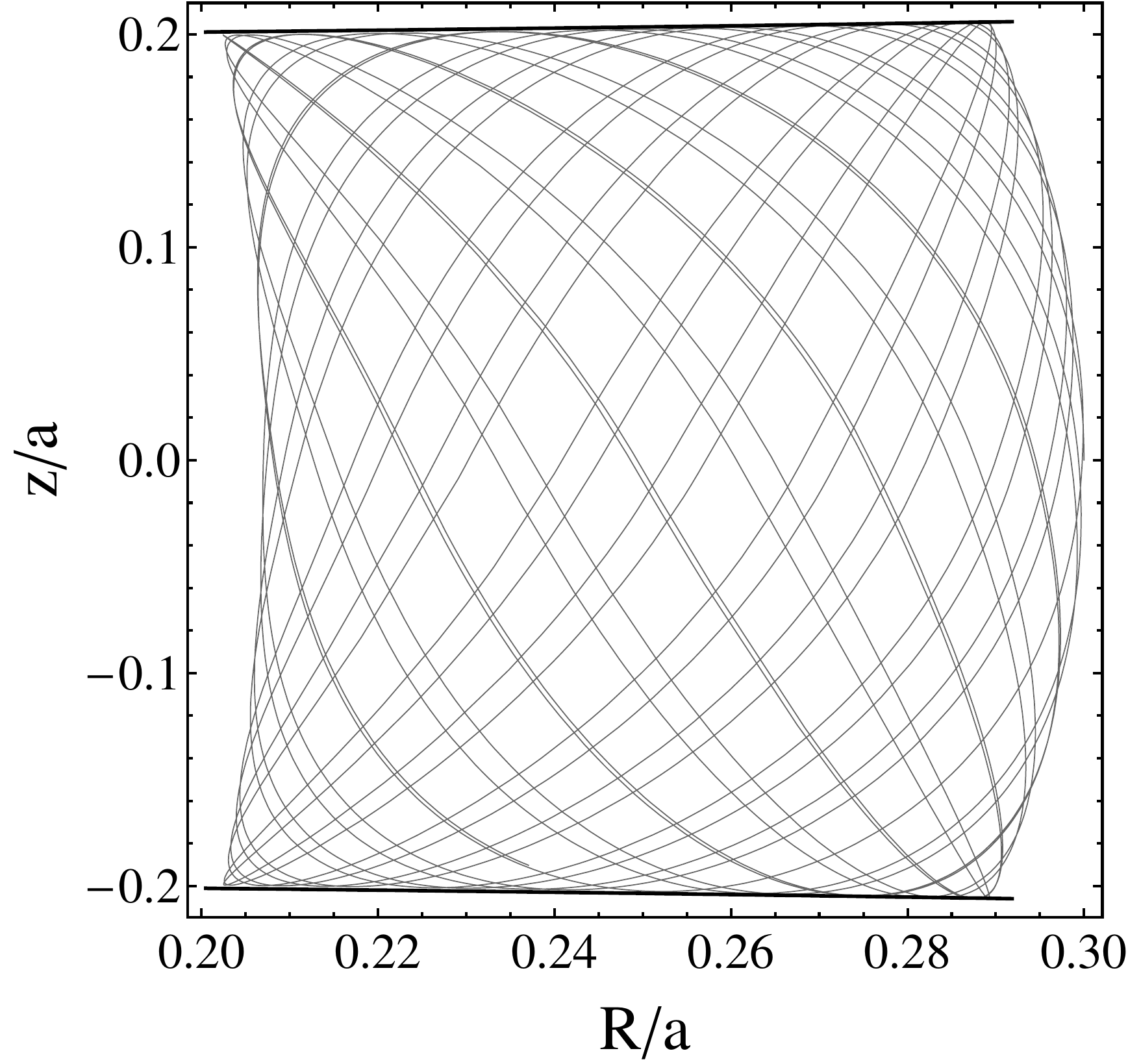,width=0.48\columnwidth ,angle=0}\quad
\epsfig{figure=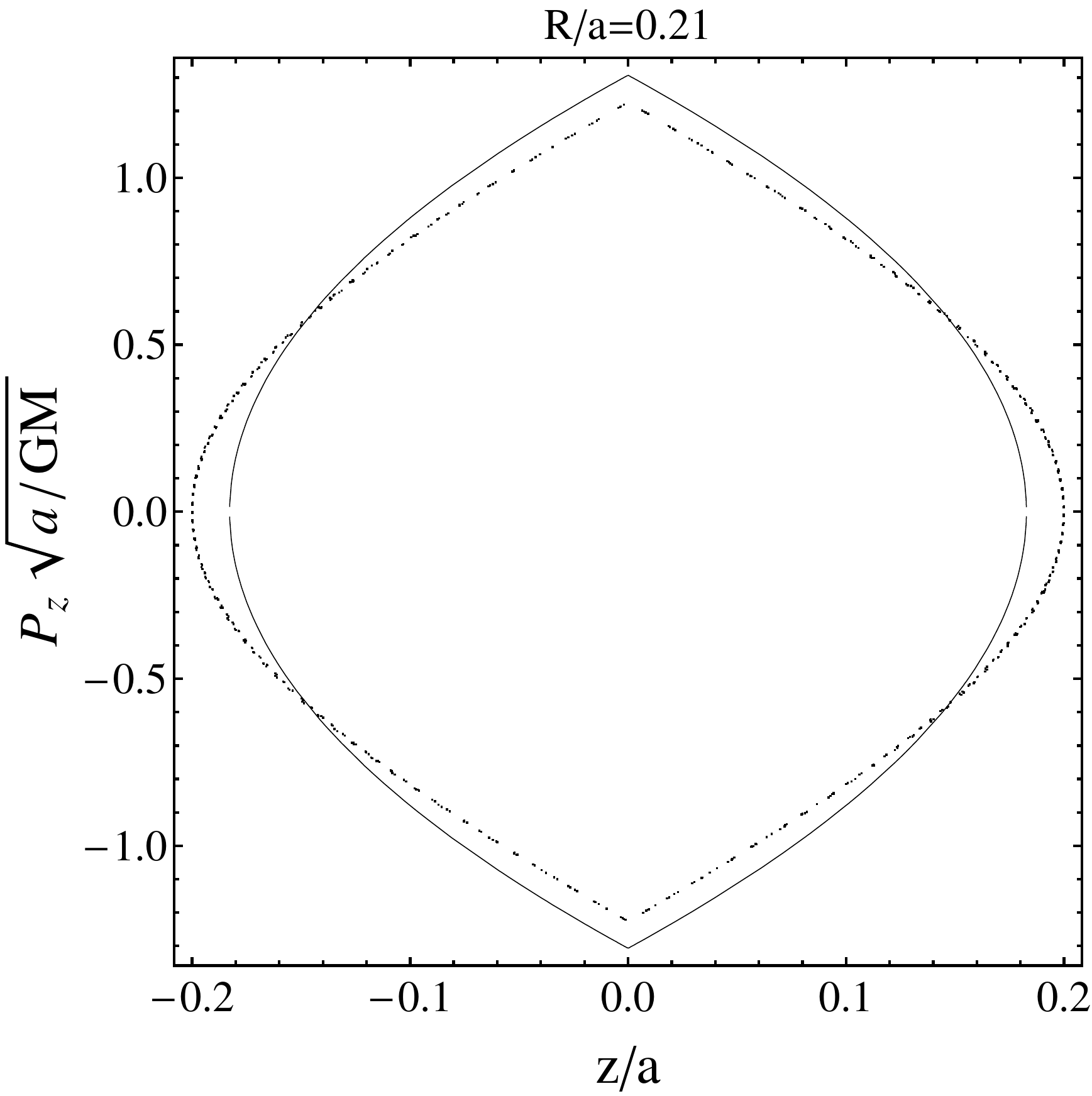,width=0.48\columnwidth ,angle=0}\\ \\
\epsfig{figure=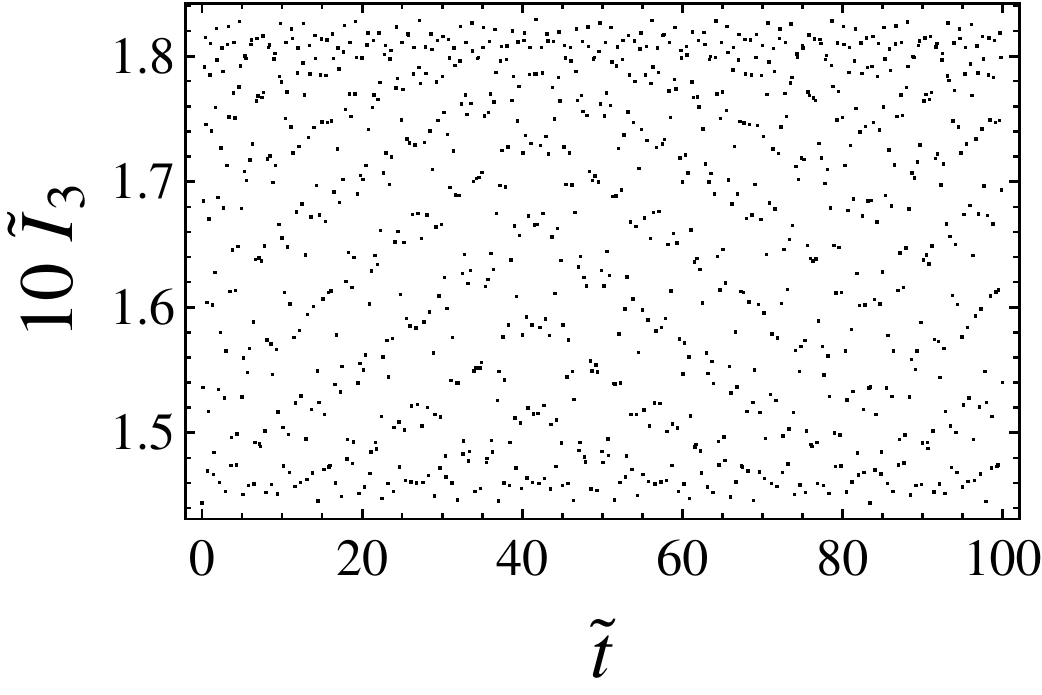,width=0.47\columnwidth ,angle=0}\quad
\epsfig{figure=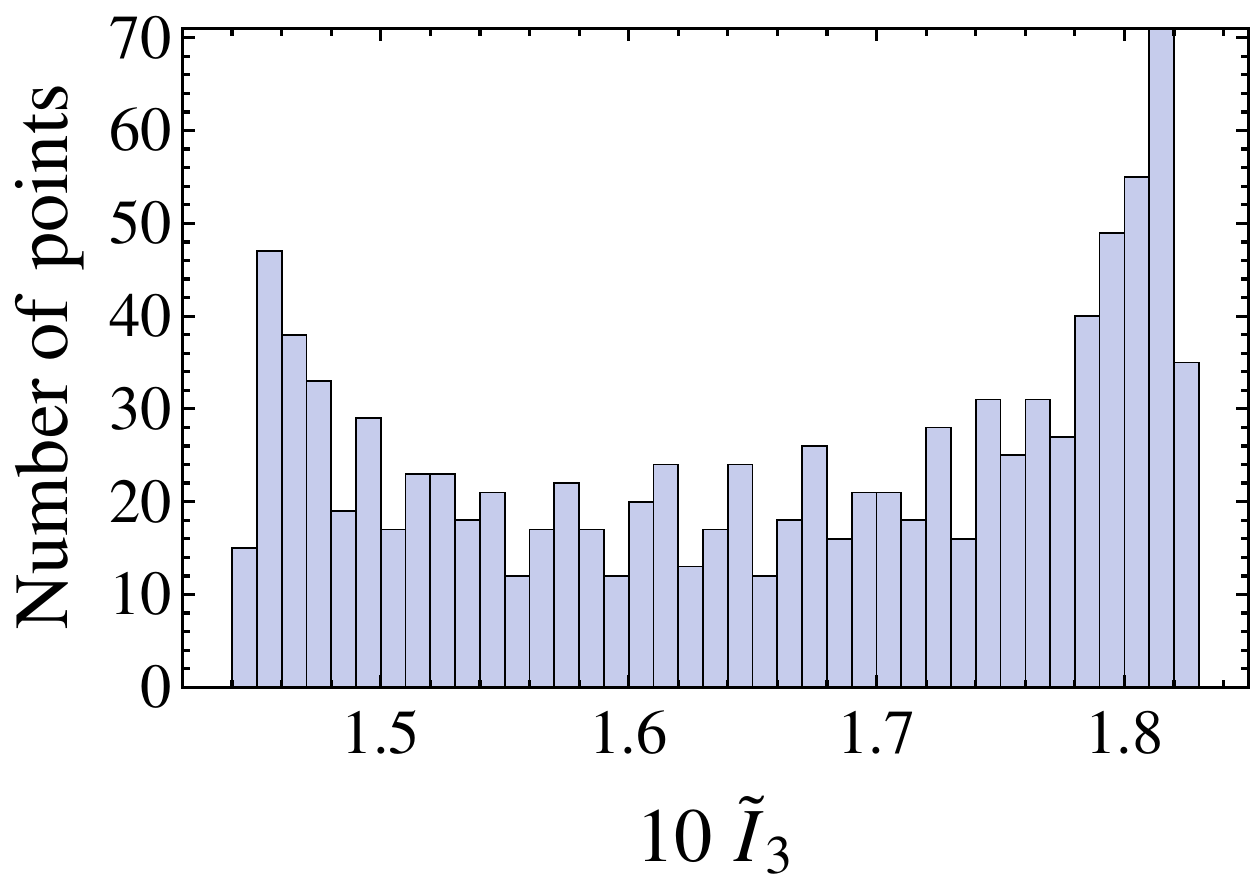,width=0.48\columnwidth ,angle=0}
\caption{Orbit in Kalnajs ($m=2$) potential with $R_{o}/a=0.3$, $P_{R}=0$ and $z_{o}/a=10^{-8}$.
The values of angular momentum and energy are $\ell/\sqrt{aGM}=0.12$, $aE/GM=-1.9$, giving $\Delta E=0.285$.
We have $T_z/T_R\approx0.91$. The Figure's style is the same as in Fig.~\ref{fig:Kuz1pp}.
The Poincar\'e section is calculated for $R/a=0.21$ and $\tilde{I}_{3,\rm mean}=0.1656$.}
\label{fig:K2b}
\end{figure}
\begin{figure}
\epsfig{figure=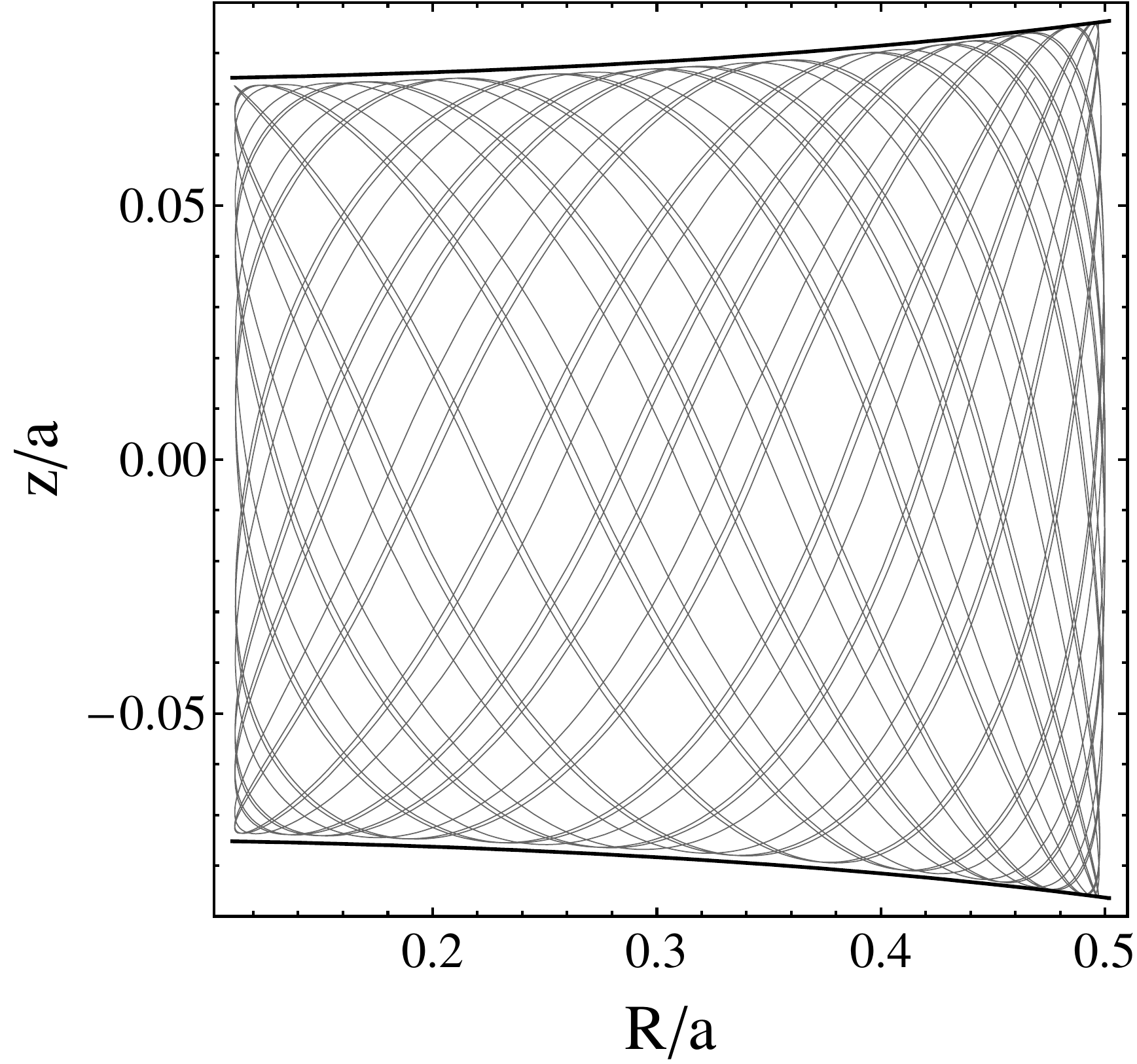,width=0.48\columnwidth ,angle=0}\quad
\epsfig{figure=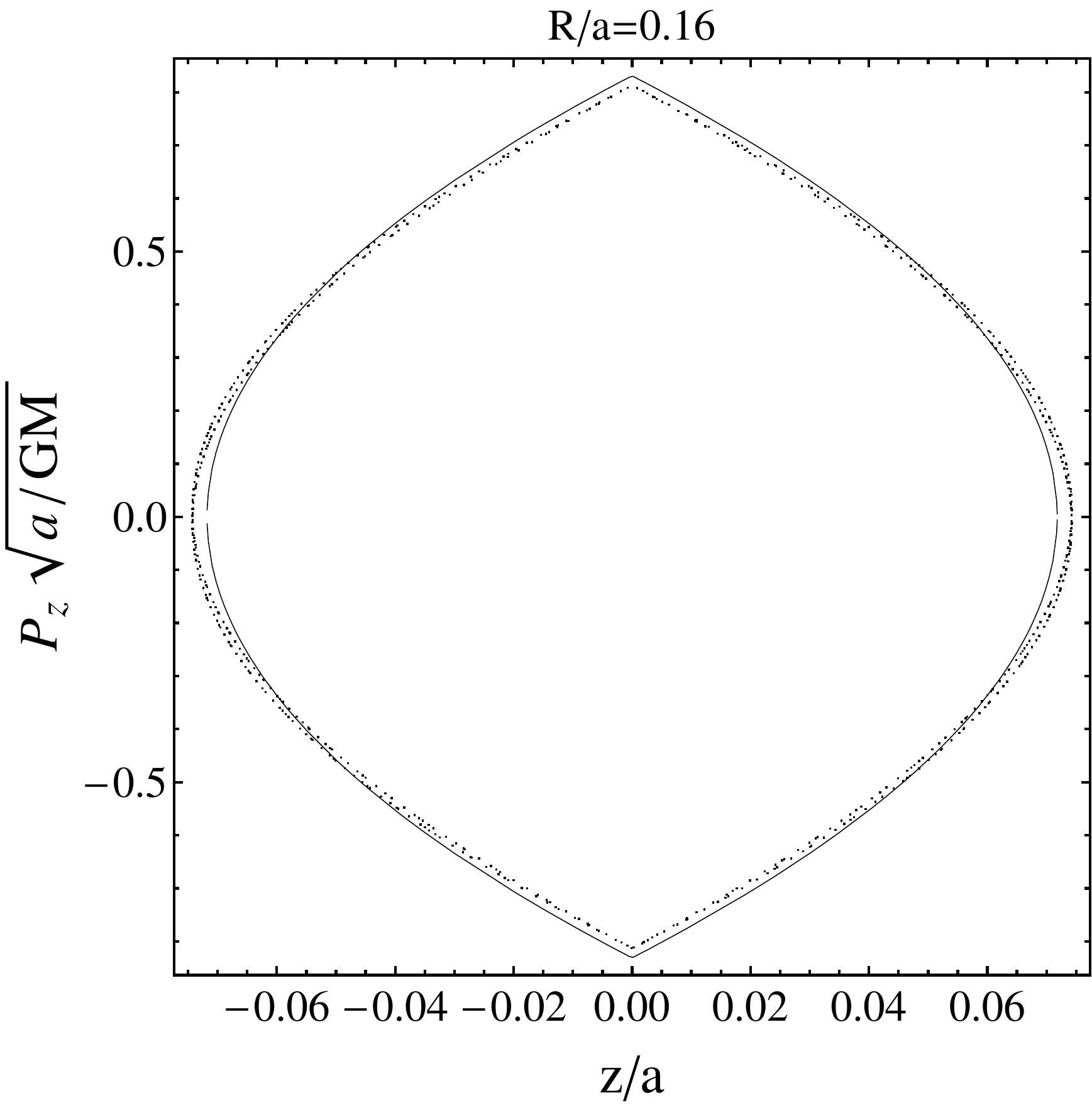,width=0.48\columnwidth ,angle=0}\\ \\
\epsfig{figure=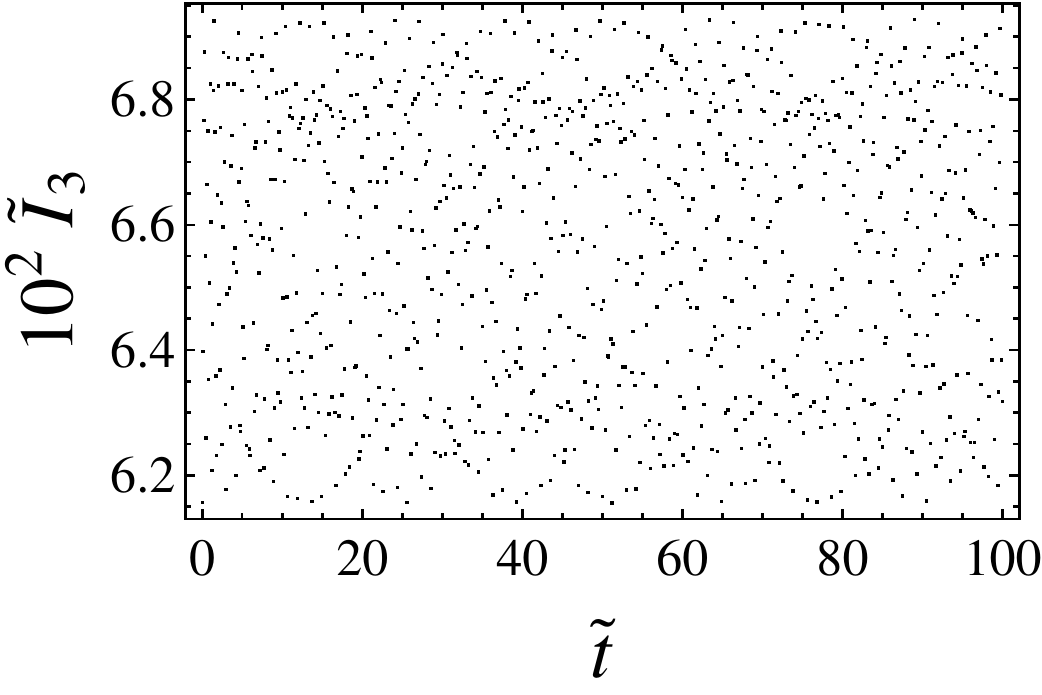,width=0.47\columnwidth ,angle=0}\quad
\epsfig{figure=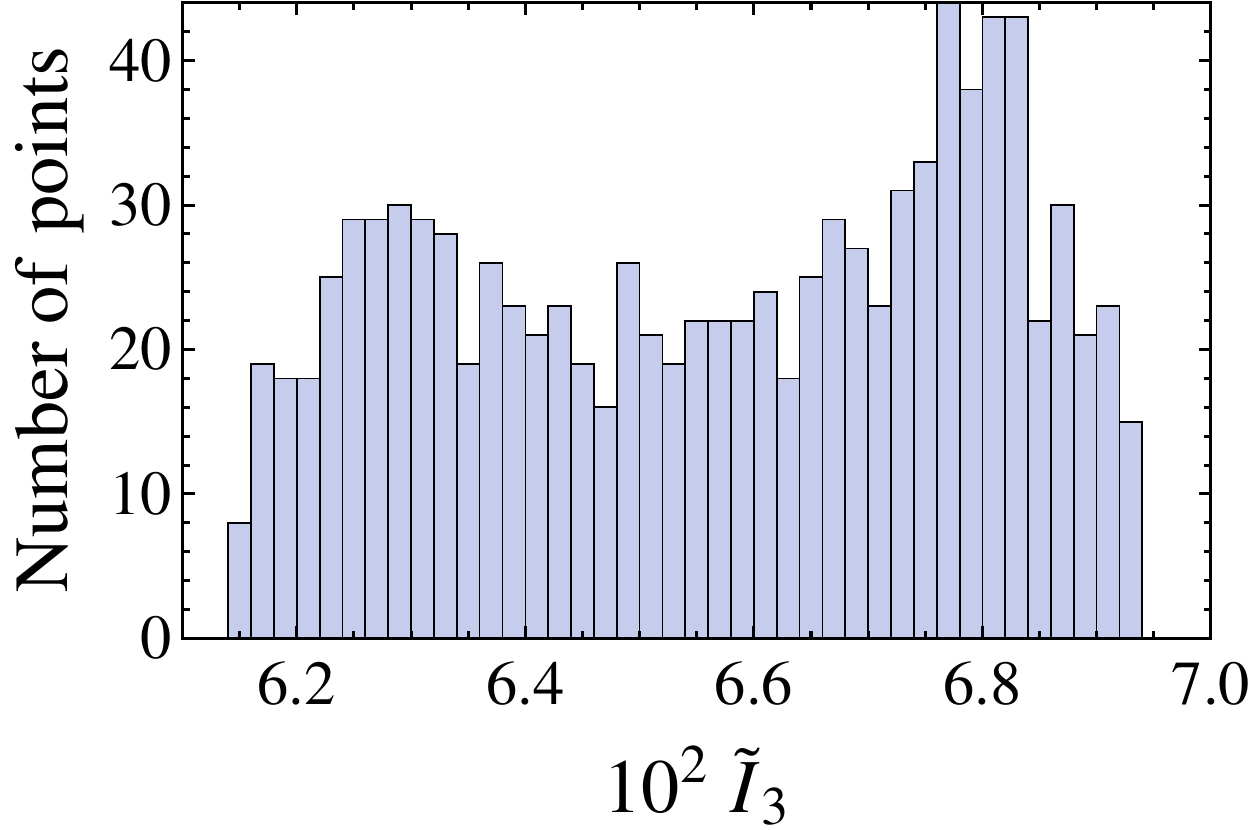,width=0.48\columnwidth ,angle=0}
\caption{Orbit in Kalnajs ($m=2$) potential with $R_{o}/a=0.5$, $P_{R}=0$ and $z_{o}/a=10^{-8}$.
The values of angular momentum and energy are $\ell/\sqrt{aGM}=0.12$, $aE/GM=-2$, giving $\Delta E=0.285$.
We have $T_z/T_R\approx0.54$. The Figure's style is the same as in Fig.~\ref{fig:Kuz1pp}.
The Poincar\'e section is calculated for $R/a=0.16$ and $\tilde{I}_{3,\rm mean}=6.656\times 10^{-2}$.}
\label{fig:K2c}
\end{figure}

As a second example, we choose the second member of the generalized Kalnajs disks \citep{gonzalez-reina2006MNRAS}, whose
surface mass density is a monotonically decreasing function of radius,
 \begin{equation}
\Sigma = \frac{5M}{2\pi a^{2}}\left( 1 - \frac{R^{2}}{a^{2}}\right)^{3/2},\label{DensKalnajs}
 \end{equation}
where $a$ is the radius of the disk  and $M$ is its total mass. The corresponding
gravitational potential can be cast in oblate spheroidal coordinates, $\xi=a^{-1}\mbox{Re}[\sqrt{R^{2}+(z-\mbox{i}a)^{2}}]$,
$\eta=-a^{-1}\mbox{Im}[\sqrt{R^{2}+(z-\mbox{i}a)^{2}}]$, through the relation
\begin{eqnarray}
\Phi_{K2} &=& - \frac{GM}{a} \left[ \cot^{-1} \xi + A
(3\eta^{2} - 1)\right. \nonumber \\
 & & \left.\quad \quad \quad \quad \quad \quad\quad + B ( 35 \eta^{4} - 30 \eta^{2} + 3)\right], \label{eq:4.23}
\end{eqnarray}
with
\begin{subequations}\begin{align}
A &= \frac{5}{14} \left[(3\xi^{2} + 1) \cot^{-1} \xi - 3 \xi \right], \\
B &= \frac{3}{448} \left[ (35 \xi^{4} + 30 \xi^{2} + 3) \cot^{-1} \xi - 35 \xi^{3} -
\frac{55}{3} \xi \right].
\end{align}\end{subequations}
The potential level contours and force field are shown in Fig.~\ref{fig:Pot-force} bottom.
This potential leads to a rotation curve which has a maximum inside the disk, in contrast with
the first member of the family (the well-known Kalnajs disk), which describes a configuration rotating as a rigid body.

It is known that motion in this potential presents
chaotic behavior \citep{ramoscaro-lopezsuspes-gonzalez2008MNRAS}. Numerical experiments  show that the range of applicability of
(\ref{eq:ZZ'sigma})--(\ref{eq:I3integral}) in this potential is much smaller than in the Kuzmin disk. The predictions for $Z(R)$ and $I_3$ remain valid for $\Delta E\ll1$, corresponding
to orbits with very small radial and vertical amplitudes (Fig. \ref{fig:K2a}). For low angular momentum (such that the orbit
is confined to the inner regions of the disk), orbits with small radial span may have higher vertical amplitudes and still satisfy Eqs.~(\ref{eq:ZZ'sigma})--(\ref{eq:I3integral}) (see Fig.~\ref{fig:K2b}). However, once the radial span of the orbit is increased, equation (\ref{eq:ZZ'sigma})
is not sufficient anymore
to describe the orbits' envelopes (see Fig. \ref{fig:K2c}, where the orbit has the
same value of $\Delta E$ as in Fig. \ref{fig:K2b}). As the angular momentum increases
(still keeping the orbits in the disk region $R<a$), the range of validity of the prediction from adiabatic invariance
decreases considerably, being valid only for orbits with small enough vertical amplitudes.
All orbits of Figs.~\ref{fig:K2a}--\ref{fig:K2c} have $T_z/T_R<1$.


\subsection{Orbits in Kuzmin disk $+$ Plummer halo}\label{sec:numericalKuzPlum}

\begin{figure}
\epsfig{figure=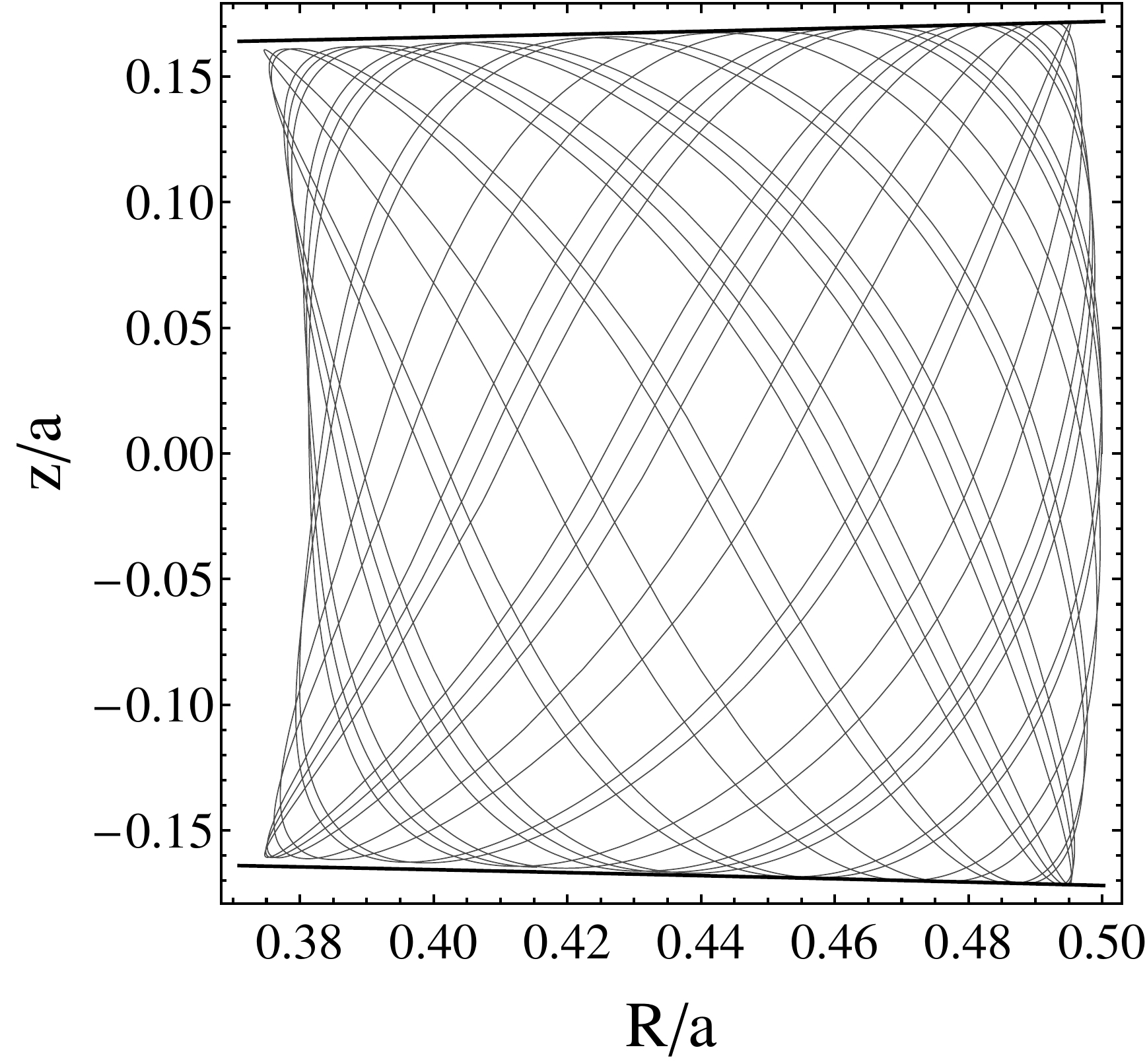,width=0.48\columnwidth ,angle=0}\quad
\epsfig{figure=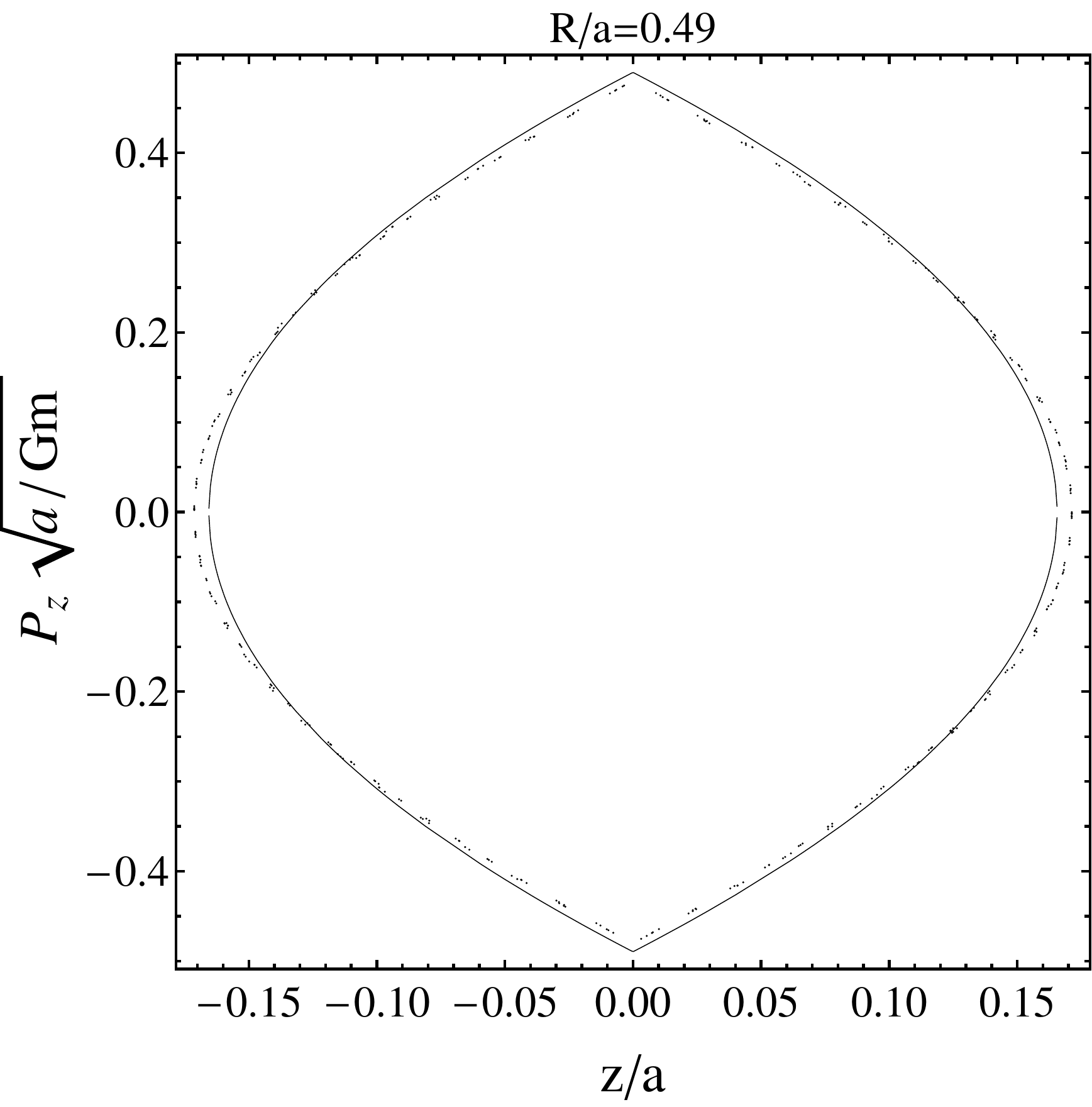,width=0.48\columnwidth ,angle=0}\\ \\
\epsfig{figure=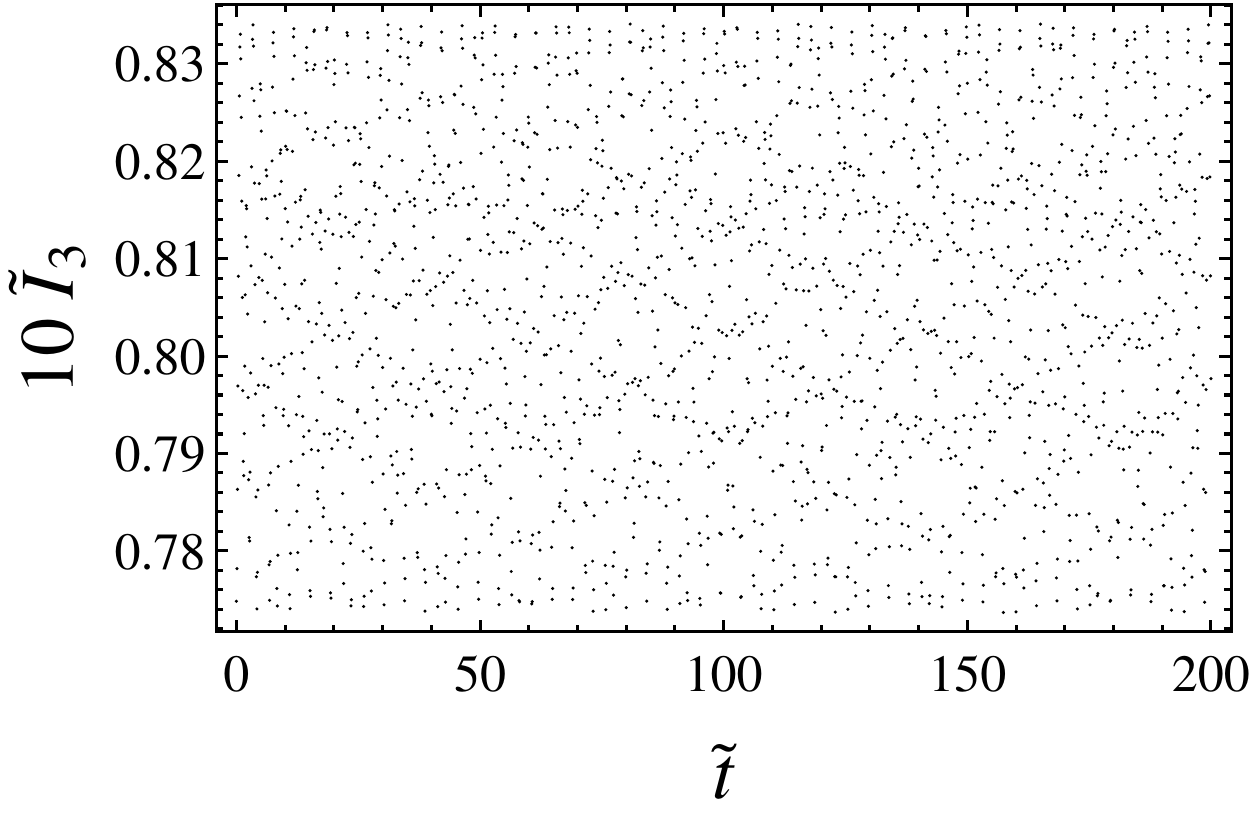,width=0.47\columnwidth ,angle=0}\quad
\epsfig{figure=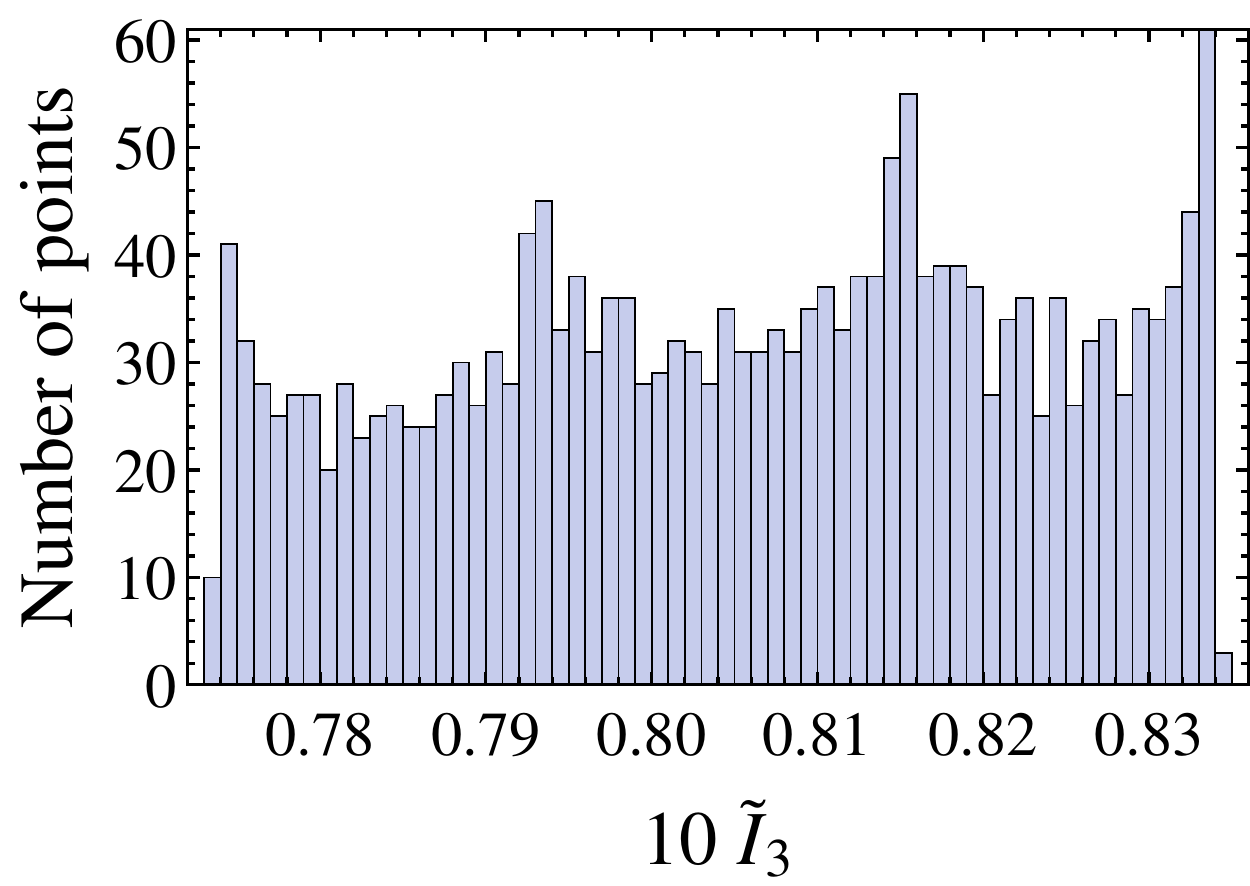,width=0.48\columnwidth ,angle=0}
\caption{Orbit  with initial conditions $R_{o}/a=0.5$, $P_{R}=0$ and $z_{o}/a=10^{-9}$, in a potential due to the superposition
of a Kuzmin disk and a Plummer halo such that $M/m=0.05$ and $b/a=1.5\times10^{-2}$.
The values of angular momentum and energy, as well as the initial conditions, are the same
as in Fig.~\ref{fig:Kuz1peq}: $\ell/\sqrt{aGm}=0.2$, $aE/Gm=-0.8$, giving in this case $\Delta E=1.38\times 10^{-1}$ and
 $T_z/T_R\approx0.65$. The initial conditions and orbit's parameters are the same as the parameters of Fig.~\ref{fig:Kuz1peq}; however, the presence of the halo increases the orbit's vertical amplitude by one order of magnitude.
The Poincar\'e section is also calculated for $R/a=0.49$.
We see that the prediction of Eq.~(\ref{eq:I3integral}) is a good approximation for this value of vertical amplitude and $\Delta E$.
The spread in $I_3(t)$ grows, in relation with fig. \ref{fig:Kuz1peq}, and the consequents in the Poincar\'e section
deviate a little from the curve predicted by  $\tilde{I}_{3,\rm mean}=8.057\times 10^{-2}$.}
\label{fig:KuzPlum1peq}
\end{figure}

\begin{figure}
\epsfig{figure=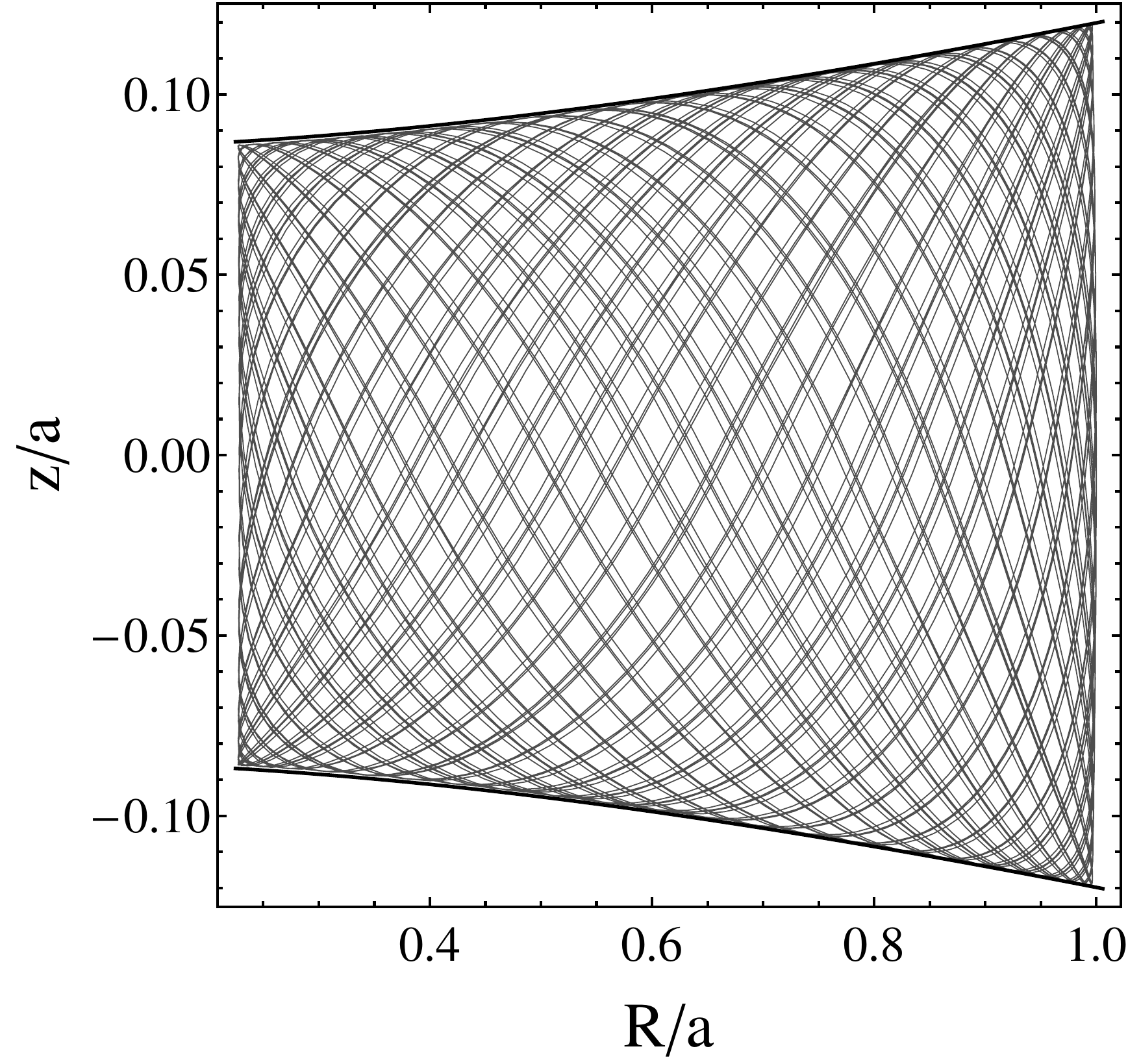,width=0.48\columnwidth ,angle=0}\quad
\epsfig{figure=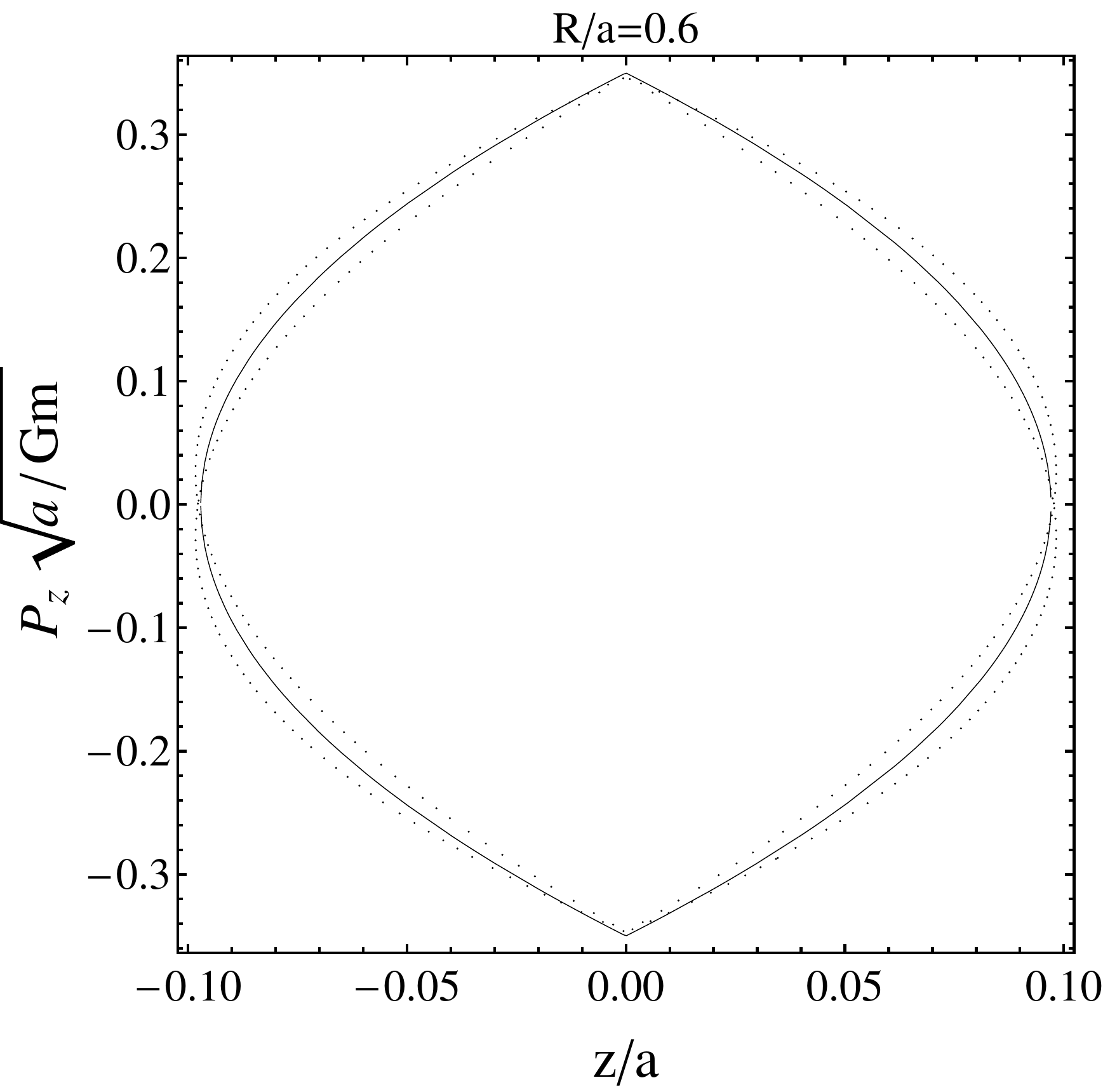,width=0.48\columnwidth ,angle=0}\\ \\
\epsfig{figure=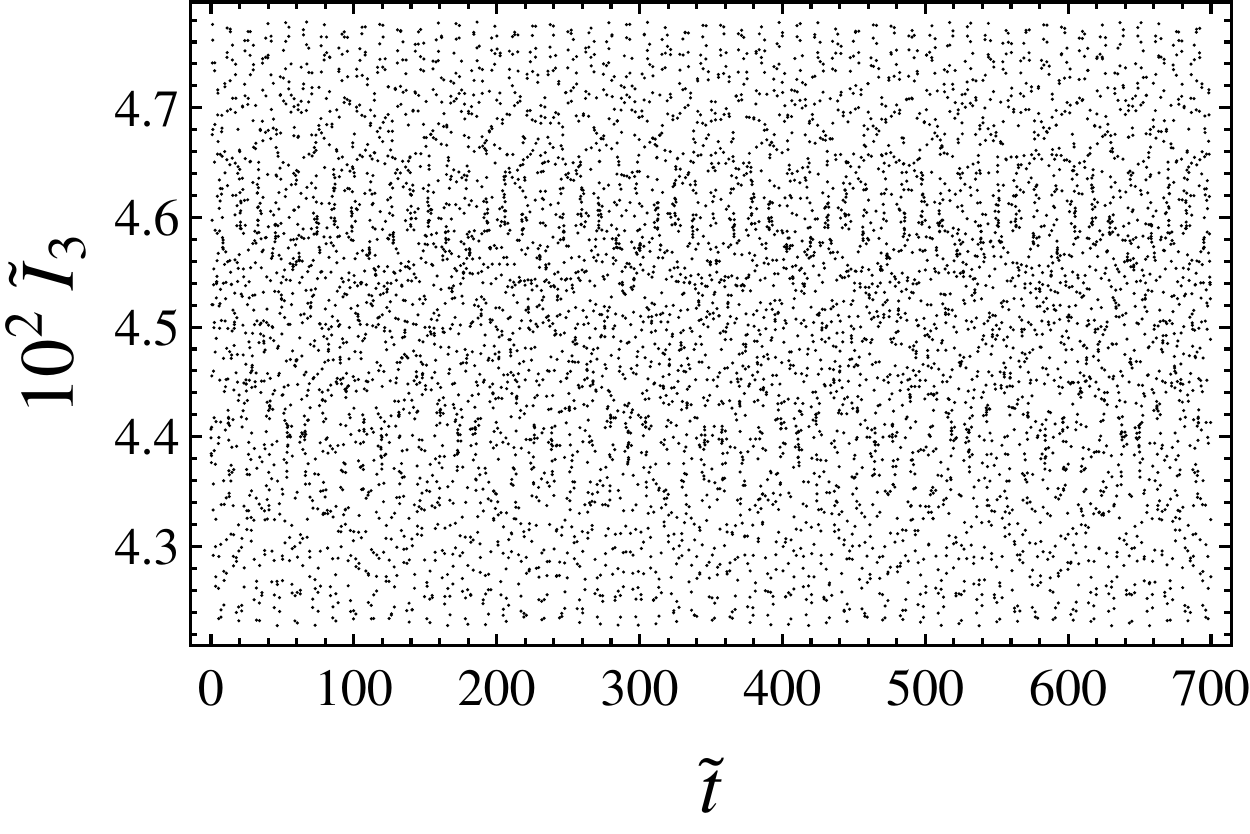,width=0.47\columnwidth ,angle=0}\quad
\epsfig{figure=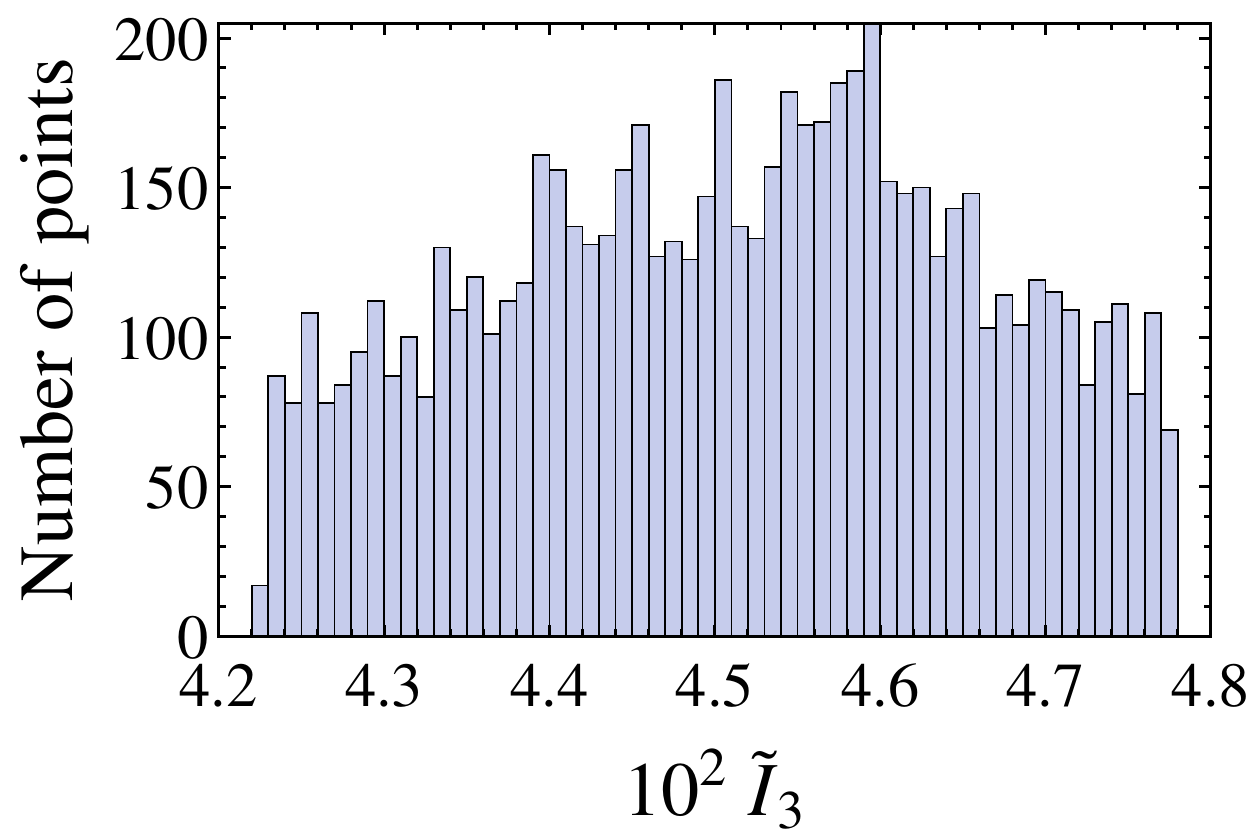,width=0.48\columnwidth ,angle=0}
\caption{Orbit  in the Kuzmin + Plummer potential, with initial conditions
$R_{o}/a=1.0$, \mbox{$P_{R}=0$}, $z_{o}/a=10^{-9}$, $\ell/\sqrt{aGm}=0.2$, $aE/Gm=-1.0$, and system's parameters $M/m=0.5$ and $b/a=1$, giving $\Delta E=2.125\times 10^{-1}$ and
 $T_z/T_R\approx0.61$. The Poincar\'e section is also calculated for $R/a=0.6$.
We see that the prediction of Eq.~(\ref{eq:I3integral}) remains a good approximation
 for this value of vertical amplitude and $\Delta E$, even when the ratio $M/m$ is one order of magnitude greater than in Fig.~\ref{fig:KuzPlum1peq}.
The spread in $I_3(t)$ grows, in relation with Fig.~\ref{fig:KuzPlum1peq}, however the consequents in the Poincar\'e section
remain close to the curve predicted by  $\tilde{I}_{3,\rm mean}=4.512\times 10^{-2}$.}
\label{fig:KuzPlum2}
\end{figure}

Now we consider the case corresponding to the superposition
of a Kuzmin disk (eq. (\ref{PotDensKuz})) and a Plummer halo,
whose potential-density pair is given by \citep{binney-tremaine:GD}
\begin{equation}
   \Phi_p = - \frac{GM}{\sqrt{r^2 + b^2}},\qquad
   \rho_p = \frac{3M}{4\pi b^3}\left(1+\frac{r^2}{b^2}\right)^{-5/2},
  \end{equation}
where $r^2 = R^2 + z^2$. We find a variety of orbits (e.g. Fig.~\ref{fig:KuzPlum1peq}), with sufficiently low energy, whose envelopes are well described by
relation (\ref{eq:ZZ'sigma}). This was obtained for different combinations of the parameters $M/m$ and $b/a$,
even for the case in which the mass of the Plummer sphere
is not negligible, as in Fig.~\ref{fig:KuzPlum2}, where $M/m=0.5$
(lower ratios $M/m$ give a wider region of validity for equation (\ref{eq:ZZ'sigma})).
However, as the halo mass grows, the range of validity of the prediction described above decreases (Fig.~\ref{fig:KuzPlum2}): deviations from
(\ref{eq:ZZ'sigma}) and (\ref{eq:I3integral}) begin to appear for lower relative energies $\Delta E$, especially in regions where the density of the sphere
is higher, $R\lesssim b$. The spherical contribution has a
negative impact on the range of validity of the predictions for $Z(R)$ and $I_3$, what may be explained by the fact that (\ref{eq:ZZ'sigma})--(\ref{eq:I3integral})
take	 into account only the razor-thin disk contribution.


\section{Conclusions}\label{sec:conclusions}

We provide an approach to address the problem of obtaining simple analytical
expressions for the  third integral of motion, associated with a great variety of disk-crossing
orbits in galactic models incorporating a razor-thin disk.
The approximated third integral of motion is given by Eq.~(\ref{eq:I3integral}) with the orbits' envelopes given by (\ref{eq:ZZ'sigma}), and, in general, this is valid for
orbits with small vertical amplitude. However,  numerical experiments  also reveal that
the quantity $I_3$ (see Eq.~(\ref{eq:I3integral})) is approximately conserved for a variety of orbits with larger values of $Z$
(i.e. of the order of the system's characteristic radius).
The same behavior was observed in models where the stellar distribution is represented by a thickened disk
\citep[see][]{vieira-ramoscaro2014ApJ}.
It is worth pointing out that Eq.~(\ref{eq:I3integral}) is also valid in the presence of a 3D 
distribution surrounding the razor-thin disk, such as a thick disk or a halo. However, as the mass of this 3D-component grows, 
the predictions of (\ref{eq:I3integral}) tend to be less accurate, since it only takes into account the potential of the razor-thin component.

A fundamental step in obtaining formula (\ref{eq:I3integral}) was the
study on the stability of equatorial circular orbits under small vertical perturbations, in models incorporating an
equatorial discontinuity of the gravitational field.
We find that, in  every galactic razor-thin disk
model,  all the equatorial circular orbits inside the matter distribution are vertically stable.
It  leads us to point out that all circular orbits of the thin-disk models presented in
\citet{gonzalez-reina2006MNRAS,letelier2007MNRAS,vogt-letelier2009MNRAS,
gonzalez-plataplata-ramoscaro2010MNRAS,ramoscaro-lopezsuspes-gonzalez2008MNRAS,ramoscaro-pedraza-letelier2011MNRAS,pedraza-ramoscaro-gonzalez2008MNRAS}
are stable, contrary to the statements made in such references.

We obtained the formula described above via a first-order approximation in $|z|$ to the vertical
 dependence of the effective potential, in the vicinity of the equatorial plane.
A better analytical description of off-equatorial orbits in systems containing razor-thin disks
would be obtained if we extended this approximate third invariant in a power series of $x=R-R_o$ and $|z|$. This procedure would be the razor-thin
version of the smooth-case Contopoulos' third integral \citep{contopoulos1960ZA, contopoulos1963AJ}.
However, such a procedure is far from being a direct extension of the first-order case we derived
here and is beyond the scope of this work. The subtleties of the $z=0$ discontinuity do not allow us to
determine the expansions of $H$ and $I_3$ in a clear, straightforward manner, since the terms involving
$\sum_{m,k}x^m |z|^k$ (k odd) and $\sum_{j}x^{N-2j} |z|^{2j}$ (with $m+k=N-1$) would be comparable. The corresponding arbitrary-order expansion is under
investigation, and will be the subject of a forthcoming paper.

The existence of a third integral of motion has a central role in the formulation of
self-consistent galactic models, in which the distribution function (DF), solution to
the collisionless Boltzmann equation, is known. As a consequence of Jeans theorem, any equilibrium DF
can be written as a function of the integrals of motion
\citep{binney-tremaine:GD}, i.e. an expression of the form $f(E,\ell, I_{3})$,
where $I_{3}$ is the non-classical third integral.
Therefore, one interesting consequence of
Eq.~(\ref{eq:I3integral}) (which works very well for
orbits with small vertical amplitude)
is that a DF of the form $f(E,\ell, I_3)$  could provide a satisfactory statistical representation
for flattened 3D self-gravitating configurations \citep{vieira-ramoscaro2014ApJ}.
Another interesting consequence of relation (\ref{eq:I3integral}) is the possibility of having at hand an alternative (or complementary)
method to measure the mass distribution of the galactic disk, in this case, from  observations determining the
vertical amplitudes of disk-crossing orbits.
This procedure may be applicable, for instance, to orbits belonging
to the thick-disk component of a given galaxy mass model, whose thin-disk component is
modeled as having zero thickness. It is also applicable to orbits in other three-dimensional distributions surrounding a razor-thin disk, e.g. the Kuzmin + Plummer system studied in sec.~\ref{sec:numericalKuzPlum}.

\section*{Acknowledgements}
The work of  RSSV is supported by the ``S\~{a}o Paulo Research Foundation'' (FAPESP), grants 2010/00487-9 and 2015/10577-9.
RSSV thanks Alberto Saa for fruitful discussions and Sylvio Ferraz-Mello for helpful remarks on the accepted version of the manuscript. The authors thank the anonymous referees for insightful comments which helped us improve the final version of the manuscript.
The authors also acknowledge the support from FAPESP grant 2013/09357-9.




\bibliography{stabN}
\bibliographystyle{CeMDAstyle}

\end{document}